\documentclass{aa}  

\usepackage{graphicx}
\usepackage{txfonts}
\usepackage{tabularx}       

\newcommand{\degree}{\ensuremath{^\circ}}

\begin{document}

\title{Influence of aerosols, clouds, and sunglint 
  on polarization spectra of Earthshine}

\titlerunning{Influence of aerosols, clouds, and sunglint on polarization spectra of Earthshine}
\authorrunning{C. Emde et al.}

\author{Claudia Emde\inst{1}
  \and
  Robert Buras-Schnell\inst{2}
  \and
  Michael Sterzik\inst{3}
  \and
  Stefano Bagnulo\inst{4}
}

\institute{Meteorological Institute, Ludwig-Maximilians-University,
  Theresienstr. 37, D-80333 Munich, Germany\\
  \email{claudia.emde@lmu.de}
  \and
  Schnell Algorithms, Am Erdäpfelgarten 1, D-82205 Gilching, Germany
  \and
  European Southern Observatory, Karl-Schwarzschild-Str. 2, D-85748 Garching, Germany
  \and
  Armagh Observatory \& Planetarium, College Hill, Armagh BT61 9DG, UK
}

\date{Received 24 October 2016; Accepted 23 April 2017}


\abstract
{Ground-based observations of the Earthshine, i.e., the light scattered by
Earth to the Moon, and then reflected back to Earth, simulate space
observations of our planet and represent a powerful benchmark for the
studies of Earth-like planets. Earthshine spectra are strongly
linearly polarized, owing to scattering by molecules and small particles
in the atmosphere of the Earth and surface reflection, and may allow
us to measure global atmospheric and surface properties of planet
Earth.}
{We aim to interpret already published
spectropolarimetric observations of the Earthshine by comparing them with
new radiative transfer model simulations including a fully realistic
three-dimensional (3D) surface-atmosphere model for planet Earth.}
{We used the highly advanced Monte Carlo radiative transfer model
MYSTIC to simulate polarized radiative transfer in the 
atmosphere of the Earth without approximations regarding the geometry,
taking into account the polarization from surface reflection and
multiple scattering by molecules, aerosol particles,  cloud
droplets, and ice crystals.}
{We have shown that Earth spectropolarimetry is highly sensitive to
all these input parameters, and we have presented simulations of a
fully realistic Earth atmosphere-surface model including
3D cloud fields and two-dimensional (2D) surface property
maps. Our modeling results show that scattering in high ice water
clouds and reflection from the ocean surface are crucial to explain
the continuum polarization at longer wavelengths as has been reported in
Earthshine observations taken at the Very Large Telescope in 2011 (3.8\,\% and 6.6\,\% at
800~nm, depending on which part of Earth was visible from the Moon at
the time of the observations).  We found that the relatively high
degree of polarization of 6.6\,\% can be attributed to light reflected
by the ocean surface in the sunglint region.  High ice-water clouds
reduce the amount of absorption in the O$_2$A band and thus explain the weak O$_2$A band feature in the observations. }
{}


\maketitle
%

\section{Introduction}

More than 3500 exoplanets orbiting stars other than our Sun have been
discovered so far and the quest to find life elsewhere in the
universe has already started.  The atmospheres of giant exoplanets are
being scrutinized for their composition and in particular for
possible thermochemical disequilibrium constituents
\citep{stevenson2010}. 
One of the greatest technical challenges is to
overcome the enormous flux difference between an overwhelmingly
bright star and the reflected light of a spatially unresolved planet.
As a contrast enhancing technique, (spectro)polarimetry exploits the
fact that the light of a solar type of star is unpolarized when
integrated over the stellar disk, while it becomes polarized by
molecular or particle scattering in the atmosphere of a planet or when it
is reflected at its surface.  Polarimetric signals provide a wealth of
information about the atmosphere and surface of a planet. 
Molecular scattering produces the largest degree of
polarization at a phase angle of 90\degree\ 
. Liquid droplets produce a particular phase angle dependence; for example, water clouds
in the atmosphere of the Earth produce a large polarization feature at phase
angles around 140\degree\ corresponding to the rainbow.  The
polarized radiance spectrum includes additional information,
for instance about trace gas concentrations of the atmosphere of a planet.  Hence
polarimetry is, in principle, an excellent tool for the detection and
characterization of exoplanets \citep{seager2000,schmidt2006}.

Up to date, Earth is the only astronomical object that can be
investigated in terms of bio-signatures unique to a life-hosting
planet and can serve as the one and only benchmark for life as we
know it.  Observations of Earthshine allow us to observe these
signatures through the reflected light of whole Earth from ground.  
Earthshine is the sunlight scattered by the dayside Earth and
reflected back to Earth from the darker portion of the visible Moon.
Different surface areas of the Earth can be probed as the relative
Sun-Earth-Moon viewing geometry changes with the phase-angle.

Linear polarization spectra of Earthshine were obtained and
interpreted by \citet{sterzik2012} 
with the FORS2 instrument mounted at the ESO Very Large
Telescope located in Chile.  The useful spectral range covered a
wavelength region between 450 and 920~nm with a resolution element of
about 3~nm.  Two distinct viewing geometries of Earth were observed
near quadrature (phase-angle near 90\degree).  During the first
observing epoch (dawn on April 25 2011), Earthshine contained
contributions from the Atlantic sea, the Amazonas, and parts of Europe
and Africa.
The second observing epoch (dusk on June 9 2011) probed the Pacific
side of Earth with almost no land surfaces visible.
The interpretation of the results critically depends on detailed
comparisons with theoretical models. Using the data of the vector
radiative transfer models of \citet{stam2008}, the fractional
contribution of clouds and ocean surface can be inferred and small amounts of vegetated areas can be diagnosed.

But important features of the observed spectra remained unexplained.
While the best models qualitatively match the blue parts of the
spectrum (between 420 and 530 nm) and allow us to derive cloud and ocean
surface coverages similar to those resembled by satellite data, the
red parts of the theoretical spectra are too low and do not fit the
observations at all.  The spectral slope of the observations is in
general much flatter (and significantly higher in the red part,
between 500 and 900nm) than predicted by the models available.  For
example, the spectropolarimetric observations by \citet{sterzik2012}
show a decreasing degree of polarization from around 10\% to 5\% in
the spectral region from 500~nm to 920~nm, while the predictions of
the simulations by \citet{stam2008} are around 10\% at 500~nm, but
less than 1\% at 920~nm.  The finding of a relatively high
polarization degree of Earthshine in the red spectral region has been
corroborated by several other measurements, such as
\citet{bazzon2013}, \citet{takahashi2013}, and \citet{miles-paez2014}.
Although the observations and their analysis are challenging
(e.g., applying a correct background subtraction of the Earthshine
spectra), all observations point in the same direction and may rather
hint to limitations in the model assumptions and their prescriptions.

%

\citet{stam2008} provide results of model calculations 
including polarization spectra of various model planets. 
The dataset was not created specifically to interpret the
Earthshine observations by \citet{sterzik2012}, in particular
the correct instrument filter function was not considered and ice clouds 
in high altitudes were not included. Thus it is not surprising that these
model data do not fit the observations well and that we need a more
realistic model setup.
\citet{stam2008} obtain the Stokes vector of light scattered by
planetary atmosphere as follows: The planet is treated as a scattering
particle and its scattering phase matrix is computed for all phase
angles simultaneously with series expansions. To calculate the elements
of the planetary scattering matrix, these authors applied the algorithm by \citet{stam2006}
based on the accurate doubling-and-adding radiative transfer code
by \citet{dehaan1987}.
This method works only for homogeneous planets, i.e., when the atmosphere is
defined by 1D profiles and the surface is defined by one 
specific surface reflection function.  
To generate datasets with different
mixtures of constituents, \citet{stam2008} approximate the light
reflected by horizontally inhomogeneous planets via weighted sums
of light reflected by horizontally homogeneous planets.  In
particular, \citet{stam2008} presents simulations for pure Rayleigh
scattering atmosphere, Lambertian surfaces (including a spectral
albedo), Fresnel reflecting surfaces, and liquid water clouds.
The disk-integration method described in \citet{stam2006} can also be
applied to inhomogeneous planets. \citet{karalidi2012a} investigated
the validity of the approximation of horizontally inhomogeneous
planets by a weighted sum of homogeneous planets. These authors divided the
planet into pixels that can have different reflection properties. For
each pixel the Fourier components of the reflection matrix are
computed and from those the disk-integrated signal is derived. They
found that for
the intensity and degree of polarization the impact of
inhomogeneity is significant. 

Simulations for a planet covered with liquid and ice water clouds have been
presented by \citet{karalidi2012b}. As \citet{stam2008} has already
shown, the strong rainbow feature in the degree of polarization may be
used to detect liquid water clouds on exoplanets. Ice clouds above
liquid clouds dampen the rainbow feature, but according to the
simulations by \citet{karalidi2012b} this feature should still be sufficiently
strong to be detected in an exoplanet with Earth-like cloud cover.
As the simulations by \citet{karalidi2012b} were performed only at three
wavelengths (550~nm, 660~nm and 865~nm), 
the detailed spectral slope of the degree of polarization is unknown.

We developed a novel Monte Carlo approach that allows us to
simulate the whole planet in full spherical geometry in one radiative
transfer simulation, making a disk integration scheme redundant. This
approach was combined with an importance sampling method that allows
us to simulate the full spectrum based on photon paths calculated for
one wavelength.  We used a backward Monte Carlo approach as in
\citet{garciamunoz2015} and \citet{garciamunoz2015b},
however we used a different sampling method for the
photon directions after scattering events. 
We included planet, star, and
receiver in our simulation; the field of view of the receiver includes
the full planet, which is illuminated according to the geometry (i.e., star
and receiver positions with respect to planet). Thus we obtain the
(spatially resolved) radiance measured by the receiver without any
approximations related to the geometry; for example, we do not need to assume
a locally plane-parallel atmosphere.

We validated our approach by comparison to the spectra provided
by \citet{stam2008} and in general find a very good agreement for the
same input assumptions.  In this contribution we demonstrate with our method
that in particular scattering in high ice water clouds and reflection at the
ocean surface have a high impact on
the expected degree of polarization and could explain the
measurements. When we use a realistic Earth model, including
three-dimensional (3D) distributions of cloud ice water and liquid water
content and a two-dimensional (2D) surface albedo map, we obtain a good
match with the observations. 

The paper is organized as follows: Section~\ref{sec:montecarlo} briefly
describes our Monte Carlo code MYSTIC and our approach to simulate
polarized radiances reflected from an illuminated planet. 
For validation we show spectral simulations for a molecular
atmosphere and compare those to the data by
\citet{stam2008}. Section~\ref{sec:observation} shows observed
polarimetric Earthshine spectra. Section~\ref{sec:o2a_band} focuses on
the degree of polarization in the O$_2$A band region.  In
Section~\ref{sec:sim_atm} we evaluate the impact of typical Earth-like
aerosols on the polarized radiance spectra, and
we present a sensitivity study for liquid and ice clouds.
In Section~\ref{sec:comp_observations} we use our
findings to interpret the observations by \citet{sterzik2012}
qualitatively. Finally, Section~\ref{sec:conclusions} presents a summary
of our results and conclusions and provides an outlook.


\section{Monte Carlo approach to simulate polarization spectra of
  Earthshine}
\label{sec:montecarlo}

\subsection{Vector radiative transfer model for Earth-like planetary atmospheres}

All simulations were performed via the
radiative transfer model MYSTIC \citep[Monte Carlo code for the
phYsically correct Tracing of photons in Cloudy atmospheres; ][]{mayer2009}, which is a
versatile Monte Carlo code for atmospheric radiative transfer. This code is
operated as one of several radiative transfer solvers of the
libRadtran software package. The one-dimensional (1D) version of MYSTIC is freely
available at \url{www.libradtran.org} \citep{emde2016} and the full 3D version used here is
available for scientific use in joint projects. The MYSTIC code may be used to
calculate polarized solar and thermal radiances and also to determine
irradiances, actinic fluxes, and heating rates. The model has been applied
extensively to generate realistic synthetic polarized satellite measurements for
remote sensing of the Earth. These data have been used for the
validation of various retrieval algorithms for cloud and aerosol
optical and microphysical properties \citep[e.g.,][]{davis2013,stap2016}. 
The MYSTIC code allows the definition of arbitrarily
complex 3D clouds and aerosols, an inhomogeneous surface albedo, and
topography. 

Polarized surface reflection is also included
for ocean. It should be noted that it is not yet possible to combine
bidirectional surface reflection functions (BPDFs) and Lambertian
surface albedos into a 2D surface reflectance properties map. Further
BPDFs for land surfaces are not included yet.

The model can be operated in fully spherical geometry. The implementation of
1D spherical geometry is described in \citet{emde2007}. In order to
simulate observations of inhomogeneous exoplanets 
as observed by a camera or telescope far away from the planet, we included 3D
spherical geometry in MYSTIC following
\citet{deutschmann2011}.   

We implemented polarization with a combination of the
following two methods \citep[for details refer to][]{emde2010}. First, the local
estimate method \citep{marchuk1980, marshak2005} was adapted to
account for polarization, which is essential for accurate radiance
simulations. Second, an importance sampling method was used to sample the
photon direction after scattering or surface reflection.

We included sophisticated variance reduction methods
\citep{buras2011}, which allowed us to calculate unbiased radiances for scattering
media that are characterized by strongly peaked phase functions without
approximations, such as delta scaling or truncation of the phase
function. The variance reduction methods have been validated in a
model intercomparison study including scattering media with strongly peaked
phase functions \citep{Kokhanovsky2010}. 

The MYSTIC code has been validated against benchmark data and
in model intercomparison studies including discrete ordinate,
doubling-and-adding, and Monte Carlo approaches to solve the radiative
transfer problem. The implementation of polarization
has been in particular validated in \citet{Kokhanovsky2010} and \citet{emde2015},
where results from MYSTIC
agreed with the commonly established benchmark results
within its standard deviation. 
Of particular relevance for the simulation of polarized spectra is
the absorption lines importance sampling method (ALIS)
\citep{emde2011}, which allows the
calculation of full spectra by tracing photons at only one wavelength. The
method can be applied to
calculate broadband spectra in moderate resolution or small regions in
very high spectral resolution.
The method currently has one limitation, which is that 
when 3D cloud or aerosol fields are included, the
optical properties of the clouds or aerosols are assumed to be spectrally
constant. This assumption is fine for narrowband simulations,
for example, the O$_2$-A band region, but not for the full spectrum from 400~nm
to 1000~nm.  

The performance of our model for the simulation of polarized
Earthshine spectra is summarized in Table~\ref{tab:cpu_times}).
In the following Sections different model setups are used, for example, 1D
spherically symmetric atmospheres in combination with ALIS or 3D
inhomogeneous atmospheres without using
ALIS; these setups are summarized in Table~\ref{tab:model_settings}.
All simulations were performed using the MYSTIC solver included
in {\sl libRadtran}, version 2.0.1.
\setlength{\extrarowheight}{5pt}
\begin{table*}
  \centering
  \begin{tabularx}{\hsize}{m{0.05\hsize} m{0.08\hsize} m{0.1\hsize} m{0.1\hsize} m{0.15\hsize} m{0.15\hsize}  m{0.05\hsize} m{0.05\hsize}}
    \hline
    Section  & model\newline geometry &  Molecules & Aerosols & Clouds & Surface & ALIS & Sample \newline resolution \\ 
    \hline
    2.2 & 3D{$^1$} & US-standard{$^3$} & - & 3D ECMWF$^6$ & 2D MODIS$^7$ & - & 100$\times$100\\ 
    2.3 & 1D{$^2$} & McClatchey{$^4$} & - & - & Lambertian
    \newline or ocean-BPDF & yes & 1$\times$1 \\  
    4.1 & 1D{$^2$} & MLS{$^3$} & - & - & black surface & yes & 1$\times$1 \\
    5.1 & 1D{$^2$} & MLS{$^3$} & OPAC{$^5$} & - & Lambertian & yes & 1$\times$1 \\
    5.2 & 1D{$^2$} & MLS{$^3$} & - & 1D layer & black surface & yes & 1$\times$1 \\
    5.2 & 1D{$^2$} & MLS{$^3$} & - & 1D layer & Lambertian & yes & 1$\times$1 \\
    6.1 & 1D{$^2$} & MLS{$^3$} & - & 1D layer & Lambertian \newline or ocean-BPDF & yes &  1$\times$1 \\
    6.2.1 & 3D{$^1$} & US-standard{$^3$} & - & 3D ECMWF$^6$ & 2D MODIS$^7$ & - & 500$\times$500 \\
    6.2.1 (O$_2$A) & 3D{$^1$} & US-standard{$^3$} & - & 3D ECMWF$^6$ & 2D MODIS$^7$ & yes & 1$\times$1 \\
    6.2.2 & 3D{$^1$} & US-standard{$^3$} & - & 3D ECMWF$^6$ & ocean-BPDF & - & 100$\times$100 \\
    6.2.2 (O$_2$A) & 3D{$^1$} & US-standard{$^3$} & - & 3D ECMWF$^6$ &
    ocean-BPDF & yes & 1$\times$1 \\
    \hline
    \multicolumn{7}{l}{$^1$spherical planet with inhomogeneous atmosphere in three spatial dimensions} \\
    \multicolumn{7}{l}{$^2$spherical planet, spherically symmetric atmosphere (horizontally homogeneous, vertically inhomogeneous)} \\
    \multicolumn{7}{l}{$^3$molecular atmosphere with 51 vertical layers \citep{anderson1986}} \\
    \multicolumn{7}{l}{$^4$molecular atmosphere with 16 vertical layers, as defined in \citet{stam2008}, Table 1}\\
    \multicolumn{7}{l}{$^5$OPAC aerosol, standard 1D profiles from {\sl libRadtran} \citep{emde2016}, 13 vertical layers} \\
    \multicolumn{7}{l}{$^6$liquid and ice water cloud fields from ECMWF model, 720$\times$360$\times$20 grid cells covering the whole planet} \\
    \multicolumn{7}{l}{$^7$global MODIS surface albedo data, 720$\times$360 ground pixels, Lambertian}\\
    \hline
  \end{tabularx}
  \caption{Model setups for simulations shown in Sections 2--6.}
  \label{tab:model_settings}
\end{table*}
\setlength{\extrarowheight}{0pt}

\subsection{Simulated pictures of the Earth}
\label{sec:earth_pics}

In order to demonstrate our model setup, we simulated the image of
Earth as recorded from space by a CCD camera.
The image of the camera includes the full Earth and
central point of the image is at 0\degree\ latitude and 0\degree\
longitude. We defined the position of the Sun by the point of
intersection between the Earth surface and the connecting line between
Earth center and Sun center. The intersection point is specified by
latitude and longitude.  The latitude of the sun position in this
simulation is 0\degree, and in this case the longitude corresponds exactly
to the phase angle. We simulated longitudes (i.e., phase angles) of
0\degree, 30\degree, 60\degree, and 90\degree.  The model atmosphere
includes molecules according to the US standard atmosphere by
\citet{anderson1986}.  We included a realistic 2D Lambertian surface
albedo map derived from MODIS data \citep{schaaf2002}.  Further we
included 3D cloud data (liquid and ice water clouds) from the European Centre for Medium-Range Weather Forecasts
(ECMWF) model (integrated forecast system IFS, \url{http://www.ecmwf.int}).
We used the operational 9h forecast from 25 April
2011, 0 UTC. The data include 3D fields of
liquid water content, ice water content,
and the sub-grid cloud
cover. The spatial resolution of the data is 0.5\degree\ in latitude and 0.5\degree\
in longitude. The vertical dimension in the ECMWF model is pressure,
which we transfered to altitude using the hydrostatic equation. We
interpolated the data on 20 altitude layers (0 to 20\,km with 1\,km
resolution). Thus, in total we obtained 720$\times$360$\times$20 grid
cells including ice water content and liquid water content,
respectively. Since a grid cell  (0.5\degree\ latitude times
0.5\degree\ longitude) is more than 50$\times$50\,km$^2$ large at the
equator, it is usually not completely cloud covered or completely
cloud free. Thus, the ECMWF model provides a sub-grid cloud cover,
which is a number between 0 and 1 and gives the fraction of the grid cell
that is cloudy (1 means fully cloudy,  0 means cloud free). The cloud
water content is given in grams of water per cubic meter of air and
refers only to the cloudy part of the grid cell. 
We multiplied the ice water content and the liquid water content with the sub-grid 
cloud cover to obtain the correct water
contents in each grid cell. For simplicity we used
constant effective radii of 10~$\mu$m (typical for liquid water cloud
droplets) and 30~$\mu$m (typical for ice crystals) as the  
ECMWF data do not include cloud
particle sizes. Internally, in the ECMWF model parameterizations were
used to calculate effective radii of cloud particles.
We chose the wavelength of 550~nm, where we
expected to see the surface and also relatively strong 
polarization due to Rayleigh
scattering.  For this example we calculated an image with 100$\times$100
pixels. We used the backward Monte Carlo tracing technique
and calculate pixel by pixel
sequentially. The initial photon direction was randomly determined
within the field of view of the individual pixels.  For each pixel we
ran 10$^{6}$ photons, resulting in 10$^{10}$ photons for the full image.

In order to simulate the measured Earthshine spectra for a
horizontally inhomogeneous planet as shown here, we do
not need to calculate a spatially resolved image, but we may sample all
photons reflected or scattered by the Earth into one large pixel. In
the backward Monte Carlo method the initial photon directions are
randomly distributed within a field of view including the full planet.
In this case we needed much fewer photons to reach a good accuracy, i.e., 10$^{5}$
photons are usually sufficient to obtain a standard deviation smaller
than 1\%. Table~\ref{tab:cpu_times} includes computational times and
the corresponding accuracy for various setups, for example, a monochromatic
simulation at 550~nm 
for a fully realistic Earth model including 3D cloud fields
and a 2D surface albedo map takes 29~s on one CPU.
The images are not calculated to compare with the observation, but to
see whether the Earth model looks realistic (like an image taken by a
camera on a satellite) and to better understand the results. 

\begin{figure*}[t!]
  \vspace*{2mm}
  \begin{center}
    \includegraphics[width=.9\hsize]{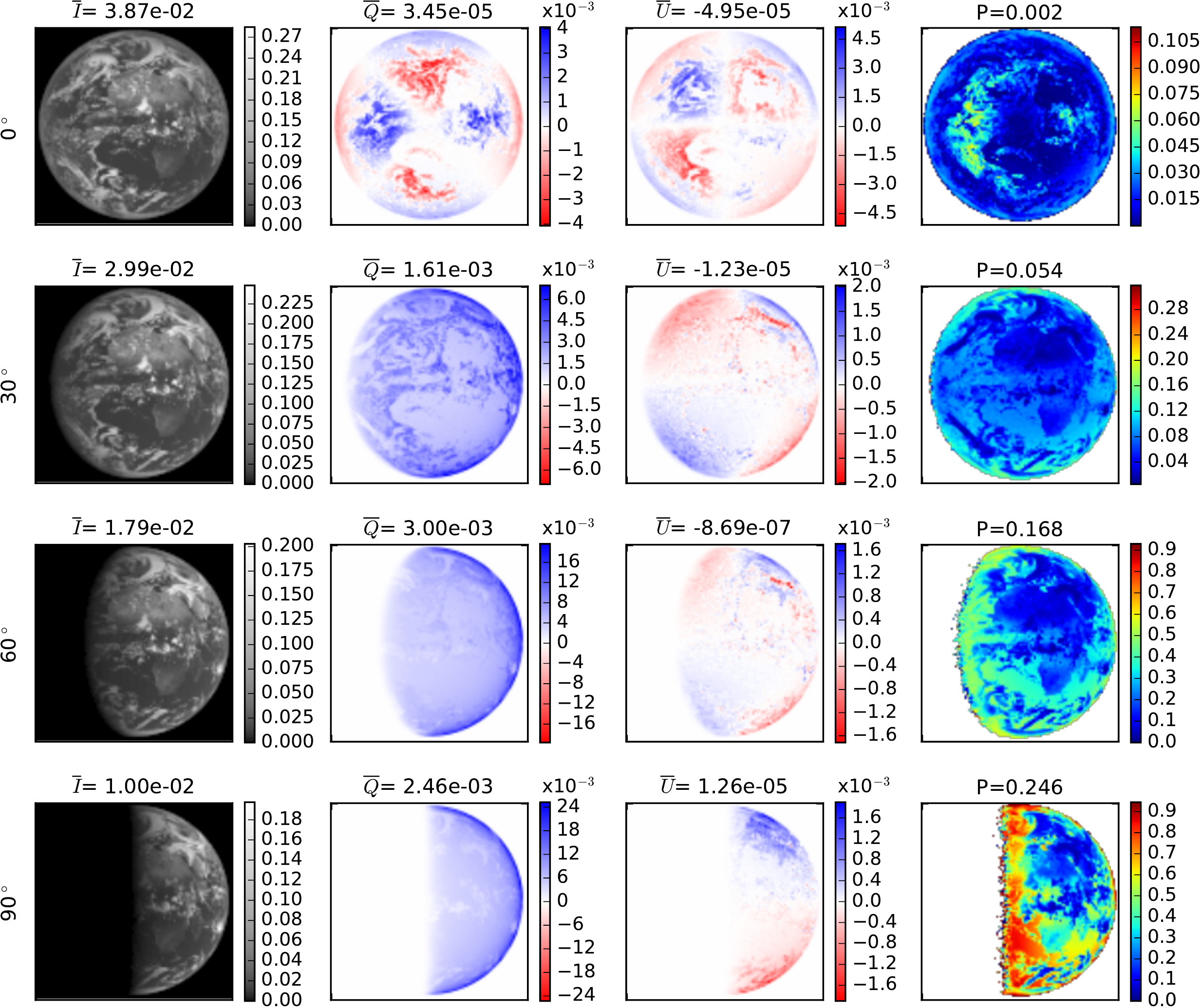}
  \end{center}
  \caption{Example simulation including a Lambertian surface albedo
    map and 3D cloud data from the ECMWF weather
    model. The rows correspond to phase angles of 0\degree, 30\degree,
    60\degree\ , and 90\degree. The columns show the Stokes vector
    components $I$, $Q$, and $U$ and the degree of linear
    polarization. On top of each Stokes component figures the mean
    values of all pixels in the image are given,  
    and the right panels include polarization calculated as 
    $P=\frac{\sqrt{\overline{Q}^2+\overline{U}^2}}{\overline{I}}$.}  
\label{fig:example1}
\end{figure*}
Figure~\ref{fig:example1} shows the results of the simulations for
images with 100$\times$100 pixels. The rows correspond to the
simulated phase angles of 0\degree, 30\degree, 60\degree\ , and
90\degree. In each row we plot the images of the Stokes vector
components $I$, $Q$ and $U$,
normalized to incoming solar
irradiance. The numbers above the images correspond to mean values
of the Stokes components, $\overline{I}=\sum{I_i}/N$, $\overline{Q}=\sum{Q_i}/N$ and
$\overline{U}=\sum{U_i}/N$, where the index $i$ denotes the pixel and 
$N=100^2$ corresponds to the number of pixels. When we perform the
simulation for one pixel only, the result corresponds exactly to
$\overline{I}, \overline{Q}, \overline{U}$ and we do not need to
average. 
The images in the last column show the degree of linear polarization ,
which is calculated as $P_i=\frac{\sqrt{Q_i^2+U_i^2}}{I_i}$. 
 We
need to calculate $P=\frac{\sqrt{\overline{Q}^2+\overline{U}^2}}{\overline{I}}$ to get the degree of polarization of the whole planet.
This can be very different from the average of the
individual values $P_i$.
 
For homogeneous planets $P$ is exactly 0 at a phase angle of
0\degree\ for symmetry reasons. 
The patterns of $Q$ and $U$ 
are symmetric with positive and negative signs, i.e., 
positive and negative polarization
values cancel each other when we observe the planet as a whole.
The inhomogeneous surface and clouds break the symmetry, therefore the degree of polarization is
nonzero. For our cloud and surface distribution, the degree of
polarization is very small ($P$=0.002). For individual pixels,
$P_i$ can be as large as 0.1. 
The degree of linear polarization $P$ increases with increasing phase
angle. We would expect the maximum at
90\degree \ for a pure molecular atmosphere over a black surface
because here we have very strong polarization due to
Rayleigh scattering. The images show that $P$ increases with
phase angle and at 90\degree\ phase angle we obtain $P$=0.246. The
phase angle dependence clearly shows the influence of molecular
scattering. 
The individual $P_i$ in clear-sky regions above ocean can be larger than
0.9.  
  
The Earth surface is clearly visible in the images of $I$ and $P$ but
only weakly visible in $Q$ and $U$ because for the simulation we assumed
that the surface is a Lambertian reflector, which by definition
reflects unpolarized radiation. The polarization by scattering of
photons that have been reflected by the surface is very small because
the photon directions after surface reflection are random.
The effects of the polarized surface reflection in the sunglint region
over ocean are discussed 
later in Section~\ref{sec:ocean_bpdf}.
The maxima of $Q_i$ and $U_i$ are at the
limb of the Earth due to second order of scattering.

We do not show the Stokes component $V$ corresponding to circular
polarization because it is very small.  
Molecular scattering does not cause  any circular
polarization and the circular polarization by 
scattering at aerosols, cloud droplets, and ice crystals 
is several orders of magnitude smaller than
linear polarization.

\subsection{Comparison to dataset by \citet{stam2008}}

To validate our approach we first calculated the same
scenarios as \citet{stam2008} for the spectral range from 300 to
1000~nm for a phase angle of 90\degree. We used the ALIS method to simulate the spectrum.
Our model yields the radiance averaged over the full quadratic field
of view, rather than the average over the planetary disk only. 
Therefore, to obtain the same normalization
as \citet{stam2008}, we need the following conversion factor:
\begin{equation}
  \label{eq:nomalization}
  N= \pi \frac{\left(2d\tan\alpha\right)^2}{\pi r_E^2} \approx \frac{4\left(\alpha d\right)^2}{r_E^2}
\end{equation}
Here $\alpha$ is the angle defining the quadratic field of view of the
simulation, $d$ is the distance between the observer and the center of
the planet, in our case the distance between Earth and Moon,
and hence $(2d\tan\alpha)^2$ is the cross-sectional area of
the field of view at the location of the Earth. The area within the
field of view covered by the planet is $\pi r_E^2$, where $r_E$ is the
radius of the Earth. The additional factor $\pi$
is required for consistency with the Stokes vector
definition by \citet{stam2008}.

\begin{figure*}[t!]
 \vspace*{2mm}
 \begin{center}
   \includegraphics[width=1.0\hsize]{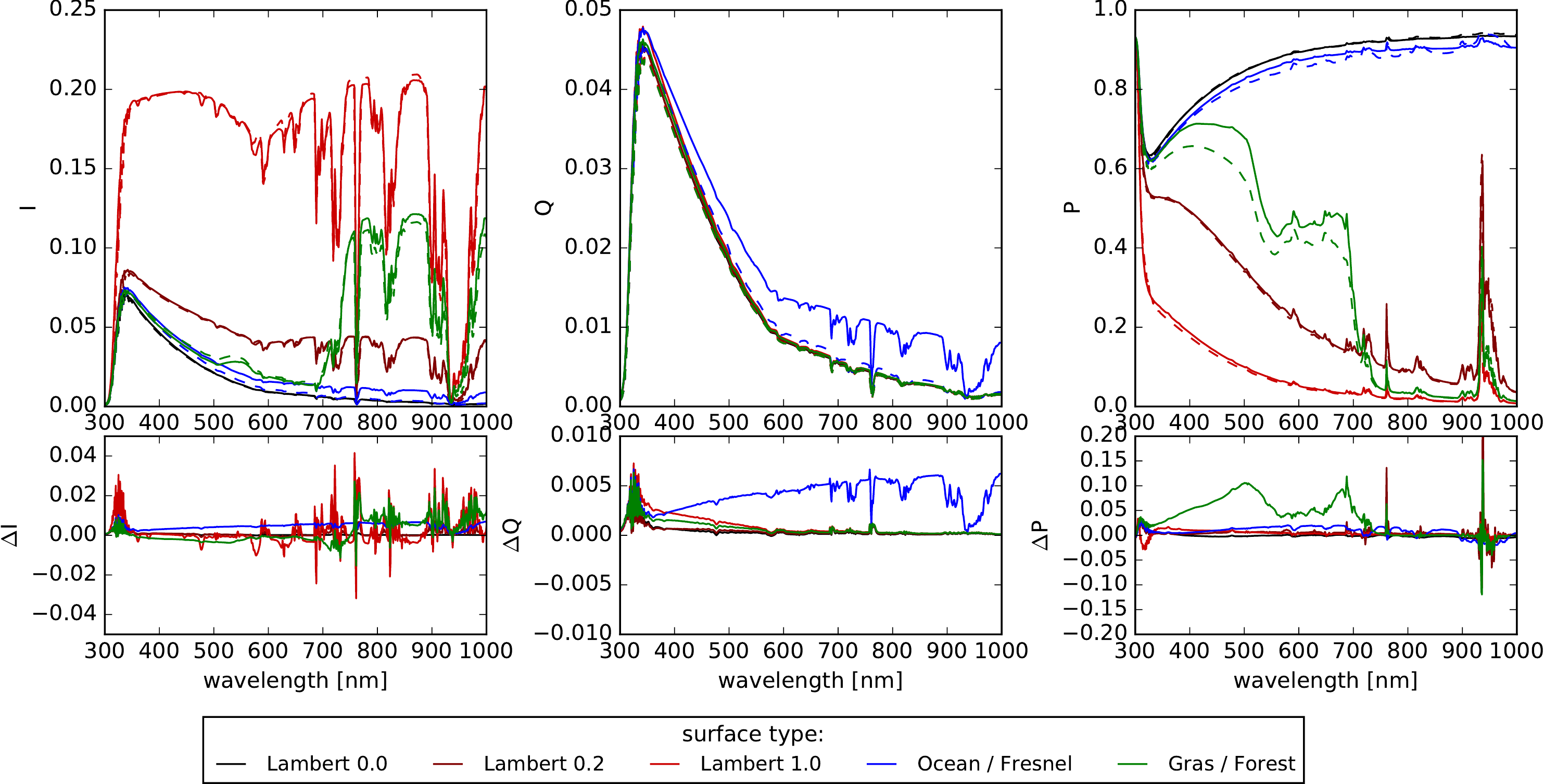}
 \end{center}  
 \caption{Intensity $I$, Stokes parameter $Q$,
   and degree of polarization $P=Q/I$ (top). The solid lines show
   MYSTIC calculations and the dashed lines show results by
   \citet{stam2008}. Lambertian surfaces with albedos 0, 0.2, and 1.0
   are compared. For ocean and land surfaces (forest and grass) the
   surface properties are not exactly the same (see text for details).
   The small bottom plots show absolute differences between MYSTIC
   results and \citet{stam2008}.}
  \label{fig:comp_stam}
\end{figure*}
Figure~\ref{fig:comp_stam} shows MYSTIC simulations in comparison to
data by \citet{stam2008}. We ran 10$^6$ photons and the relative
standard deviation is generally below 1\% for all Stokes parameters
and also for the degree of polarization. 
The bottom panels show the absolute differences between MYSTIC and the
data by Stam. 

We generally find very good agreement for Lambertian surfaces. Here we tried to adapt the scenario by
\citet{stam2008} as close as possible. We used the standard atmosphere by
\citet{mcclatchey1972} and added oxygen with a concentration of 21\%.
For absorption \citet{stam2008} has used a k distribution
\citep{stam2000}, whereas we used the REPTRAN
parameterization by \citet{gasteiger2014} in coarse resolution (15~cm$^{-1}$).
These different approaches explain the differences
in the absorption bands. 

For the land surface we used the spectral albedo of grass as measured by
\citet{Feister1995}. These measurements are available for the spectral
region from 290~nm to 800~nm. Above 800~nm we used a constant albedo
of 0.587, corresponding to the measurement at 800~nm. \citet{stam2008}
used data
for deciduous forest from the ASTER spectral library. The spectral
albedos are similar, in particular they both show a local maximum
between 500~nm and 600~nm (due to absorption bands of
chlorophyll) and both show a high albedo at wavelengths longer than
700~nm.
The difference seen in Fig.~\ref{fig:comp_stam} for land surfaces
(green lines) are due to the different spectral albedo data.

\citet{stam2008} treats the ocean as a Fresnel
surface, i.e., a flat surface neglecting the influence of oceanic
waves. For the MYSTIC simulations we used a reflection matrix that is also
based on the Fresnel equations and additionally takes
into account the influence of the waves including shadowing effects
\citep{mishchenko1997, cox54a, cox54b, tsang1985}. We chose a small
value of 1~m/s, which produces a narrow glint and should compare better
with the pure Fresnel surface than more realistic larger wind speeds.
Our model produces
larger $I$ and $Q$ values than the pure Fresnel surface; also the
degree of polarization is slightly larger. 

\section{Polarimetric observations of the Earthshine}
\label{sec:observation}

\begin{figure}[b!]
 \vspace*{2mm}
 \begin{center}
   \includegraphics[width=1.0\hsize]{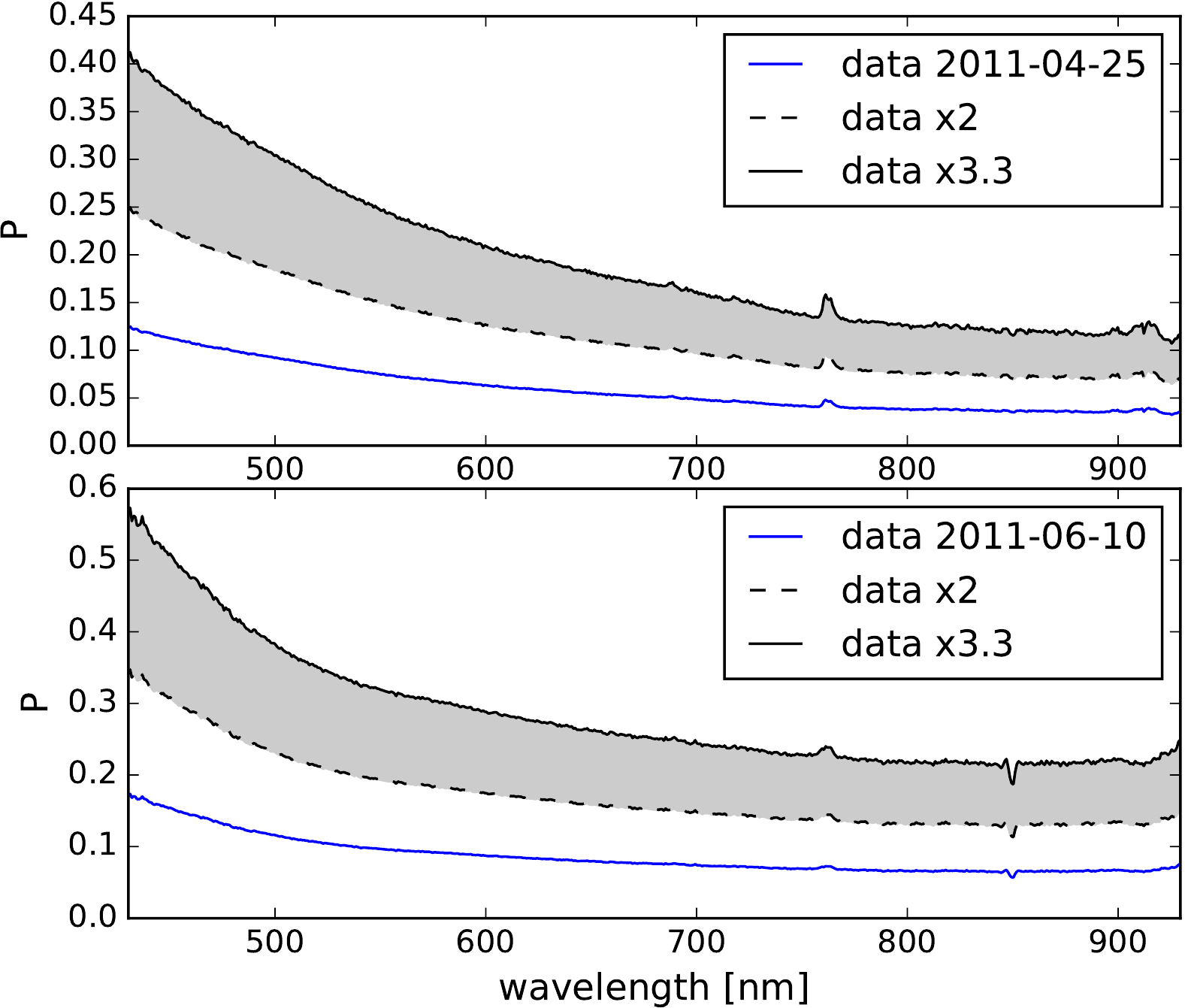}
 \end{center}  
 \caption{Observations by FORS of the degree of polarization (blue line). 
   The black lines are for different assumptions for
   moon depolarization; factor 3 (solid line) corresponds to
   Dolfuss (1957) and factor 2 (dashed line) corresponds approximately 
   to Bazzon (2013) (for
   lunar surface albedo 0.1.)}
   \label{fig:measurements}
\end{figure}
Fig.~\ref{fig:measurements} shows observed data of the FORS
instrument (blue line). The Earthshine is sunlight scattered by the
dayside Earth, which is reflected by the nightside of the Moon.  We
therefore needed to multiply the modeled Stokes vector with the lunar
depolarization matrix to compare
our model results with the Earthshine observation. Unfortunately this matrix is not known yet, and this
is indeed the largest uncertainty for the interpretation of Earthshine
measurements.
We multiplied the observed data with a lunar
depolarization factor $\delta$ to get a rough estimate of the degree
of polarization of the Earthshine, without reflection at the lunar
surface. 

\citet{dollfus1957} derived a constant value of
$\delta$=3.3, whereas
\citet{sterzik2012} assumed 
$\delta=3.3 \cdot \frac{\lambda}{550}$, where $\lambda$ is the wavelength in nm,
thus increasing depolarization
with wavelength.  \citet{bazzon2013} derived a polarization
efficiency function $\epsilon(\lambda, a)$ ($\epsilon=1/\delta$) based
on polarized reflection and albedo measurements of several Apollo lunar soil 
samples \citep{hapke1993}. The polarization
efficiency function $\epsilon(\lambda, a)$
depends on surface albedo $a$ and on
$\lambda$. For low albedos of 0.1 the polarization efficiency ranges
from 0.57 at 450~nm to 0.48 at 800~nm. For high albedos of 0.2 this value
ranges from 0.37 at 450~nm to 0.21 at 800~nm; these values correspond
roughly to the assumption by \citet{sterzik2012}. 

For the interpretation of the Earthshine observations the 
lunar depolarization factor is the largest
uncertainty. In Fig.~\ref{fig:measurements} the black lines show
the measurement multiplied with $\delta=3.3$ (solid line) and
multiplied with $\delta=2$ (dashed line), respectively.
The gray area shows the
possible range of the degree of polarization of the light scattered
by the illuminated Earth. This corresponds to the model
output, i.e., to an Earth
observation of an imaginary instrument on the lunar surface.

\section{Degree of polarization in O$_2$A-band region}
\label{sec:o2a_band}

\subsection{Importance of spectral resolution}

For the interpretation of spectral features such as the O$_2$A absorption 
band it is very important to perform the radiative transfer
simulations at the same spectral resolution as the measurements
\citep[see also][]{boesche2008}.
The FORS observations were obtained with grism 300V and a 2\arcsec\ slit
width, yielding a spectral resolution of about 250 or 3\,nm around the
O$_2$A band.

In order to demonstrate the importance of spectral resolution we took a closer
look at the O$_2$A band.
\begin{figure*}[t!]
 \vspace*{2mm}
 \begin{center}
   \includegraphics[width=1.0\hsize]{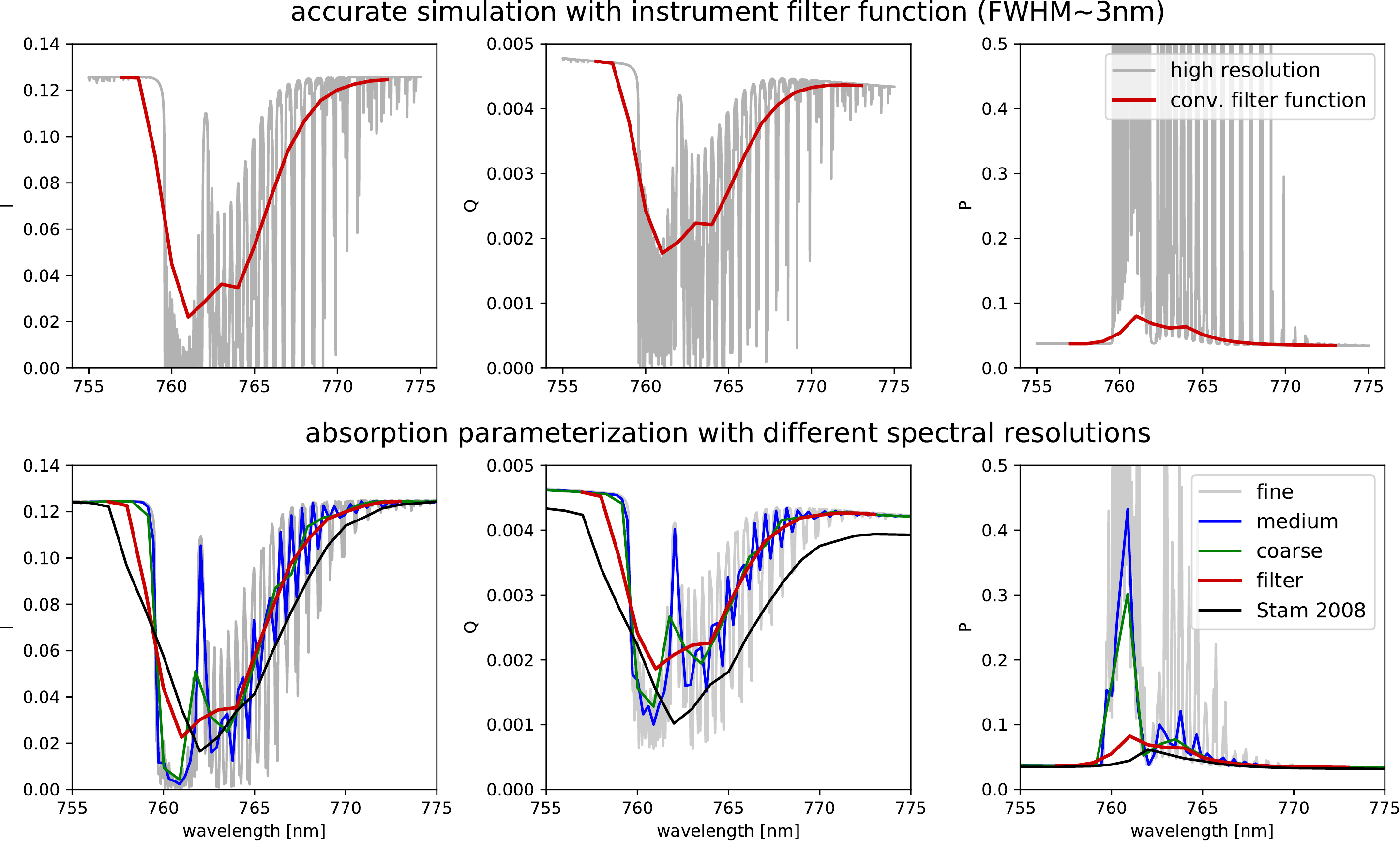}
 \end{center}  
 \caption{Intensity $I$, Stokes parameter $Q$, and degree of
   polarization $P=Q/I$ in the O$_2$A band region. The top plots show
   high spectral resolution calculations with $\Delta\lambda$=0.01~nm and
   the spectrum convolved with the instrument filter function.
   The bottom plots show the same spectral region calculated using the REPTRAN
   parameterization in three spectral resolutions (fine, medium,
and   coarse). The red line shows the REPTRAN calculation in coarse
   resolution convolved with the instrument filter function. For
   comparison, the data from \citet{stam2008} are also shown.
   The Lambertian surface albedo is 0.6.}
  \label{fig:comp_stam_o2a}
\end{figure*}
The gray lines in the top plots of Fig.~\ref{fig:comp_stam_o2a} show
a simulation
performed at a spectral resolution of 0.01~nm. The absorption
coefficients were obtained using the Atmospheric Radiative
Transfer Simulator (ARTS ) line-by-line model \citep{eriksson2011}. For this
calculation the standard midlatitude-summer atmosphere by
\citet{anderson1986} was used. The phase angle is 90\degree.

  The Stokes vector calculated in
high spectral resolution
is convolved with  the instrument filter function (red line in
Fig.~\ref{fig:comp_stam_o2a}) to obtain the correct result.
The bottom plot shows the same spectral region for different spectral
resolutions calculated via absorption parameterizations.
For REPTRAN three resolutions are available: fine (1~cm$^{-1}$ corresponding to $\approx$0.05~nm at
760~nm), medium (5~cm$^{-1}$ corresponding to $\approx$0.3~nm), and
coarse (15~cm$^{-1}$ corresponding to $\approx$1~nm). The red line in the
bottom plots shows the REPTRAN simulation in coarse resolution
convolved with the instrument filter function. The result is very
close to the accurate simulation based on the simulation in high
spectral resolution.
Results shown throughout the paper are simulated using REPTRAN in
coarse resolution and the resulting spectra are convolved with the FORS
instrument filter function. 

The data by \citet{stam2008} (black line in bottom plots)
are available on a 1~nm grid but their k-distribution method obviously
averages over a wider wavelength range and does not include
the FORS instrument filter function. 
The maximum degree of polarization depends to a large extent on the spectral resolution; this value
decreases with coarser resolution due to averaging, therefore it is
crucial to take into account the correct instrument filter function,
in particular when spectral features are compared.  

\section{Simulation of polarized Earthshine spectra for various
  atmospheric components}
\label{sec:sim_atm}

To study the effect of various atmospheric constituents on
the degree of polarization, we performed simulations separately for
aerosols, liquid water clouds, and ice water clouds.
The results of the sensitivity studies are presented in this 
section. 

A phase angle of
90\degree\ is a good observation geometry that yields high polarimetric
signals for molecular atmospheres such as the atmosphere of the Earth, thus
all simulations shown here are for this
planet-sun-observer geometry. Further we simulated the full Earth as
one pixel, which corresponds to a measurement of the Earthshine.

\subsection{Aerosols}

\begin{figure}[h]
 \vspace*{2mm}
 \begin{center}
   \includegraphics[width=1.0\hsize]{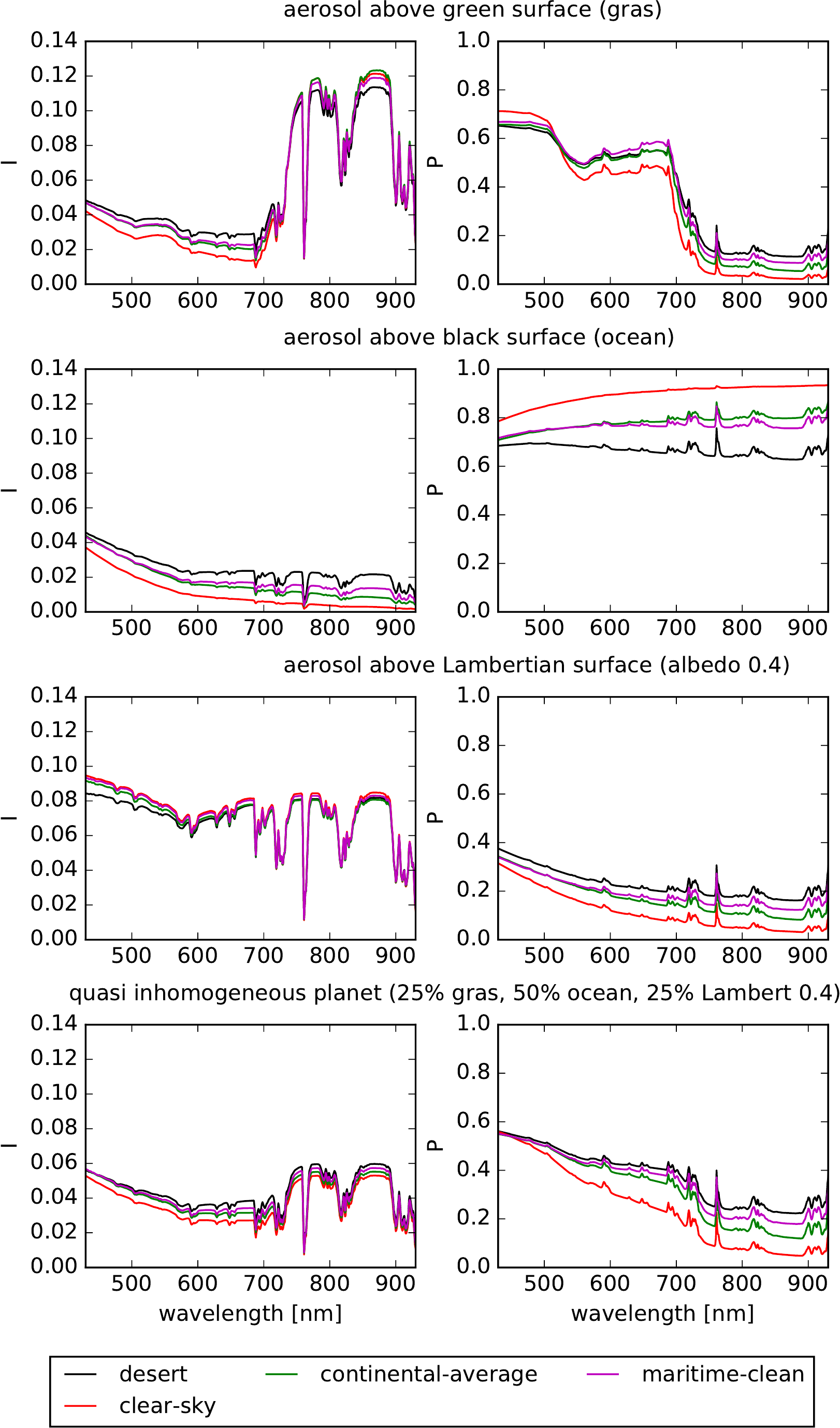}
 \end{center}  
 \caption{Intensity $I$ (left panels) 
   and degree of polarization $P$ (right panels) for an atmosphere
characterized by various aerosol
   mixtures above different surface types. From top to bottom: Grass,
   ocean, Lambertian surface with albedo 0.4, and mixed surface properties are shown.}
 \label{fig:aerosol}
\end{figure}
Figure~\ref{fig:aerosol} shows simulations for various standard aerosol mixtures
(desert, continental average, and maritime clean),
which are defined according to the Optical Properties of
Aerosols and Clouds (OPAC) database \citep{emde2016,hess1998}.
For all simulations in this and the following sections we included the
midlatitude-summer atmosphere by \citet{anderson1986}.
The top row represents a completely green planet. The polarization spectra
clearly show the vegetation step at 700~nm; this is the expected
result because typical aerosol profiles have relatively small optical
thicknesses of smaller than 0.5 and the surface is well visible. Compared
to the clear-sky simulation the degree of polarization is lower
at shorter wavelengths and higher at longer wavelengths. The reason
is that at shorter wavelengths, Rayleigh scattering is much stronger
and aerosol scattering causes smaller polarization than Rayleigh. At
longer wavelengths, Rayleigh scattering is weak and the aerosols mask
the depolarizing surface. Hence the degree of polarization becomes
larger in the presence of aerosols.
Above 750~nm the measured degree of
polarization in the continuum corresponds approximately to the
simulation with continental aerosol profile, but spectral
absorption features are much stronger in the simulation than in the
observation. 

The second row shows a simulation for a dark surface. Compared to
clear sky, aerosol scattering decreases the degree of
polarization by 10-20\%. 

The third row shows a simulation for a Lambertian surface with an
albedo of 0.4, which is similar to a sand surface. Here, the degree of
polarization is enhanced by aerosol scattering.

The fourth row shows a simulation for a ``quasi'' inhomogeneous planet. The
various simulations were 
approximated by a weighted sum as in \citet{stam2008},
\begin{equation}
  {\bf I}(\lambda, \alpha)=\sum_{n=1}^{N} f_n {\bf I}_n(\lambda, \alpha)
  \qquad {\rm with} \qquad \sum_{n=1}^{N} f_n = 1
  \label{eq:quasi_inhom}
,\end{equation}
where $N$ is the number of simulated horizontally homogeneous planets and
$f_n$ are the respective fractions. Although 
the weighted sum approach yields significantly different results
compared to simulations for inhomogeneous planets, it
can be used to estimate roughly the influence of the various components on
polarized radiance spectra.

We assumed the following fractions for the surface type:
25\% grass, 50\% ocean, and 25\% Lambertian with an albedo of 0.4. Here
also, the degree of polarization is increased by aerosol
scattering. Generally the smallest increase is observed by the
continental-average aerosol mixture, followed by the maritime-clean
aerosol. The desert aerosol mixture shows the largest impact. 
The results are dominated by the land surfaces because the total
intensity reflected by the ocean is relatively small. 

The results show that aerosols that are typically found in the atmosphere of the Earth cannot explain the observed
polarization spectra. For the simulation with surface albedo 0.4 the
degree of polarization is similar to the observations (see
Fig.~\ref{fig:measurements}), but the spectral slope is different and,
even more important, the spectral features are much weaker
in the observations than in the simulation.

\subsection{Water clouds}

\begin{figure}[t]
 \vspace*{2mm}
 \begin{center}
   \includegraphics[width=1.0\hsize]{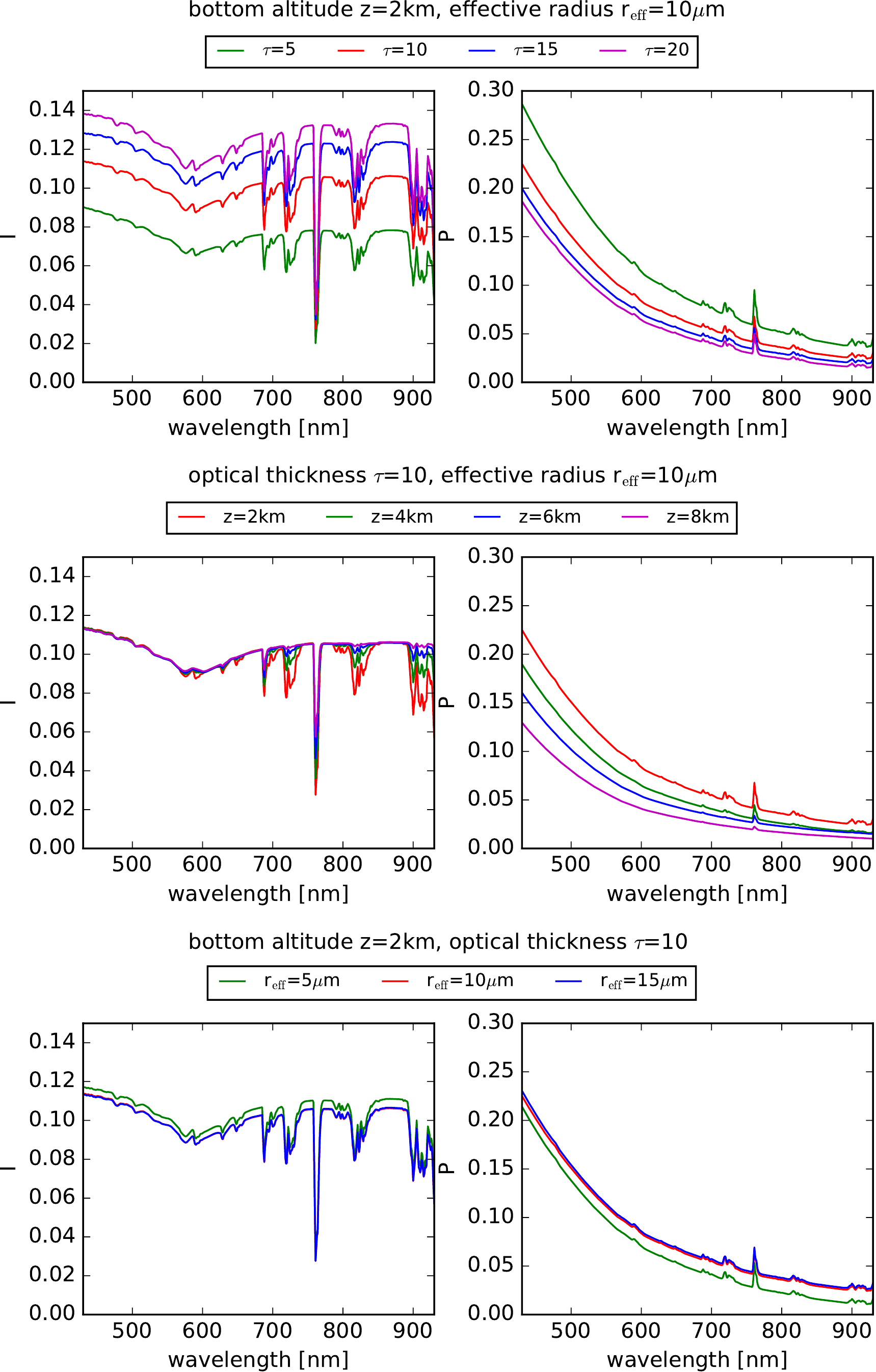}
 \end{center}  
 \caption{Intensity $I$ and degree of
polarization $P$.  The top figures show the sensitivity to cloud
optical thickness for a cloud layer from 2-3~km altitude including
cloud droplets with an effective radius $r_{\rm eff}$ of
10~$\mu$m. Cloud optical thicknesses $\tau$ at 550~nm range from 5 to 20. The
middle row shows the sensitivity to cloud altitude for a cloud with
$\tau$=10 and $r_{\rm eff}$=10~$\mu$m. The bottom altitude of the 1~km
thick cloud layer varies from 2 to 8~km. The bottom figures show the
sensitivity to effective radius for a cloud layer with a bottom
altitude of 2~km and $\tau$=10. The effective radius varies from 5
to 15~$\mu$m.}
  \label{fig:water_cloud}
\end{figure}

Figure~\ref{fig:water_cloud} shows the sensitivity of the polarized radiance
spectra on various water cloud parameters, i.e., cloud optical thickness, cloud
altitude, and effective radius of cloud droplets.
The cloud optical properties for the simulations were calculated using
the Mie tool of the libRadtran package \citep{emde2016, wiscombe80a}.
Optical properties of single spherical droplets were
averaged over a gamma size distribution with a constant effective
variance of 0.1 and different effective radii.
The surface albedo is 0 and aerosols are not included in the
simulations. The impact of aerosols and surface on radiance simulations with
water clouds is relatively small
because the clouds hide the boundary layer including
most aerosols and the surface. 

The top plots show the sensitivity to cloud optical thickness at
550~nm, which
varies between typical values from 5 to 20. The cloud layer is
situated between 2 and 3~km altitude and the effective radius of the
cloud droplets is 10~$\mu$m. The intensity $I$ increases with increasing cloud
optical thickness because the thicker the cloud the more it reflects
to space. The degree of polarization is decreased as the
optical thickness increases, mainly because the highly polarizing Rayleigh
scatterings are replaced by cloud scattering which polarize less. Also
enhanced multiple scattering decreases the degree of
polarization. 

The middle plots show the sensitivity to cloud altitude. The cloud
layer has an optical thickness of 10 and an effective droplet radius
of 10~$\mu$m. The geometrical thickness of the cloud is always 1~km and
the bottom height is varied from 2 to 8~km. The intensity $I$ is
almost invariant to cloud altitude because the amount of radiation
reflected by the cloud depends mostly on optical thickness. The degree
of polarization $P$ decreases with increasing cloud altitude because
the cloud ``hides'' the molecular atmosphere below which the amount of Rayleigh scattering, which is mainly responsible for the
polarization signal, decreases. 

The bottom plots show the sensitivity to the effective radius, which is
varied from 5 to 15~$\mu$m for a cloud layer between 2 and 3~km with
an optical thickness of 10. We see that the intensity and degree of polarization are not very sensitive to cloud droplet size in
this spectral region. The reason is that in the geometrical optics
limit, which we may assume approximately 
for the given parameters, extinction is 
2 for all droplet sizes and the imaginary part of the refractive index
of liquid water is approximately 0, i.e., extinction is only caused by
scattering. The scattering phase matrix is also invariant to particle
size in geometrical optics.

Overall we see that the degree of polarization in the red part of the
spectrum is lower ($<$5\% for $\tau\ge$10) than in the observations (10-30\%).
The observed slope of the polarization spectrum
$P$ (gray lines) is also flatter than all simulations including liquid
clouds. All results are only valid for a phase angle of
90\degree, depending on phase angle the degree of polarization can be
increased or decreased compared to pure molecular scattering. We have
not investigated phase angles other than 90\degree\ since the
observations were performed close to 90\degree.


\subsection{Ice water clouds}

\begin{figure}[h!]
 \vspace*{2mm}
 \begin{center}
   \includegraphics[width=1.0\hsize]{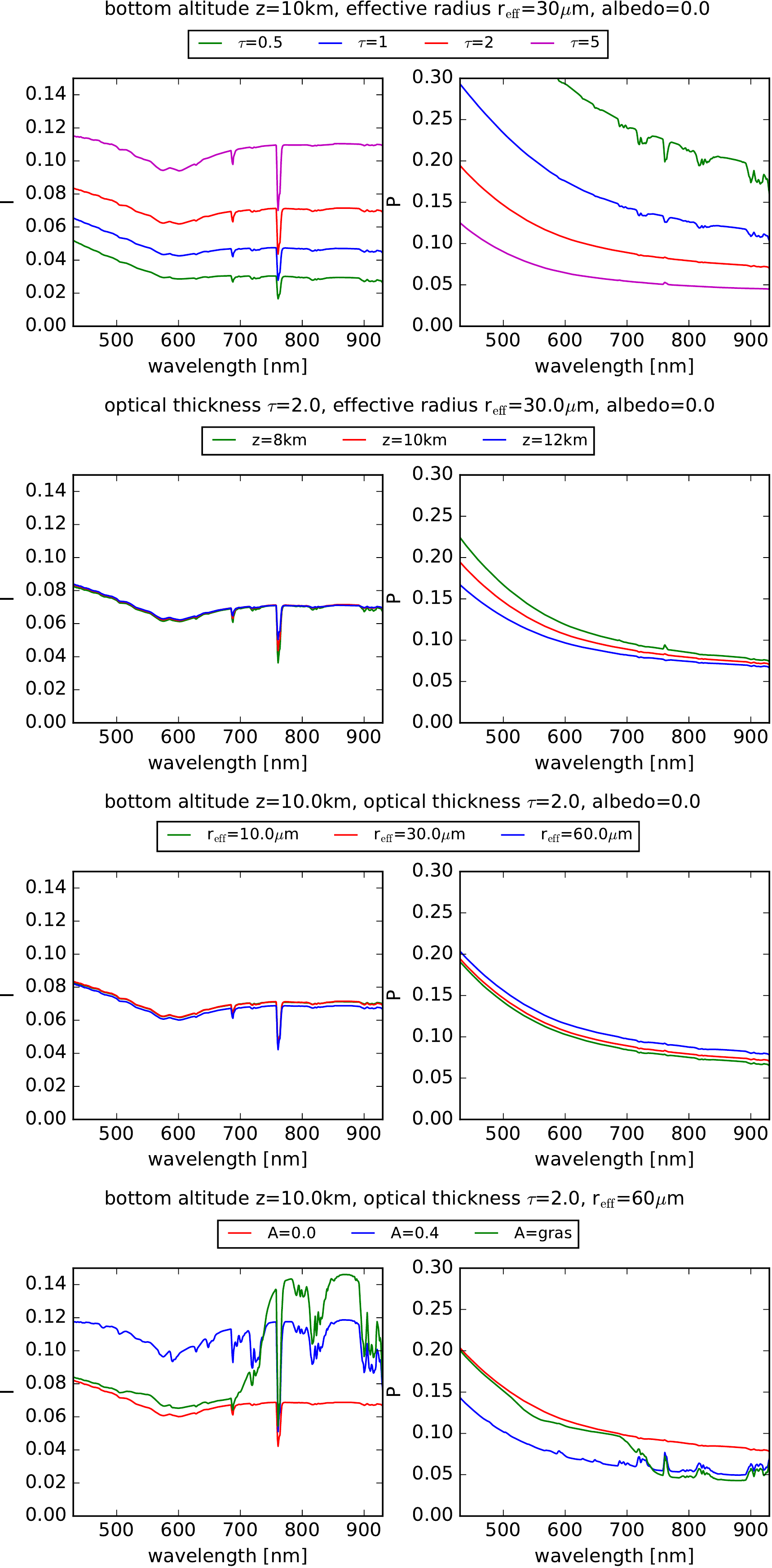}
 \end{center}  
 \caption{Intensity $I$ and degree of
polarization $P$. The top figures show the sensitivity to ice water cloud
optical thickness (at 550~nm) for a cloud layer from 10-11~km altitude including
ice crystals with an effective radius $r_{\rm eff}$ of
30~$\mu$m. Cloud optical thicknesses $\tau$ range from 0.5 to 5. 
The second row shows the sensitivity to cloud altitude for a cloud with
$\tau$=2 and $r_{\rm eff}$=30~$\mu$m. The bottom altitude of the 1 km
thick cloud layer varies from 8 to 12~km. The third row shows the
sensitivity to the effective radius for a cloud layer with a bottom
altitude of 10~km and $\tau$=2. The effective radius is varied from 10
to 60~$\mu$m. The bottom panels show the sensitivity to surface
albedo. }
\label{fig:ice_cloud}
\end{figure}
To study the sensitivity to ice water cloud parameters, we varied
the ice cloud optical thickness at 550~nm between 0.5 and 5, the cloud bottom
altitude between 8~km and 12~km, and the effective radius between
10$\mu$m and 60$\mu$m. These values are typical for cirrus clouds on
Earth; for example, \citet{wang2011} have shown that ice optical thickness derived
from MODIS data is in the range from about 0.1 and 6. The effective
diameter of the ice crystals is in the range from 20 to 120~$\mu$m.    
The mean values of the MODIS retrieval are 2.83 for optical thickness
and 58.5 for effective diameter.

Figure~\ref{fig:ice_cloud} shows the intensity $I$
and the degree of polarization $P$ for a planet covered with a 1 km
thick ice cloud layer. The ice cloud consists of ice crystals of
various habits, such as solid columns, aggregates, and bullet rosettes. For
the ice cloud optical properties the parameterizations for a general
habit mixture by
\citet{heymsfield2013}, \citet{yang2013}, and \citet{baum2014} have been
used. This mixture is composed of nine crystal shapes: solid/hollow bullet
rosettes, solid/hollow columns, plates, droxtals, small/large
aggregate of plates, and an aggregate of solid columns.

The top figures show the sensitivity to ice cloud optical
thickness. The bottom altitude of the cloud is at 10~km and the surface
albedo is 0. As for the
water cloud the intensity increases with increasing optical thickness
and the degree of polarization decreases. Further we see that the
spectral slope of $P$ decreases with increasing optical thickness. The
slope is much flatter than for water clouds because the ice cloud
layer is at a higher altitude and ``hides'' the Rayleigh scattering in
the lower atmosphere. 

For very small optical thicknesses of 0.5 and 1
we see that the degree of polarization becomes smaller in the
absorption bands than in the continuum. A possible explanation is the
following: For low
optical thickness the path length in the cloud layer is small compared
to the path lengths in the free molecular atmosphere. Therefore
absorption reduces the number  of strongly polarizing Rayleigh
scattering events more than the number of much lower polarizing cloud scattering
events.

Now we investigate how other cloud parameters influence $I$ and
$P$. The second row in Fig.~\ref{fig:ice_cloud} shows the effect of
cloud altitude;  for the water cloud $I$ is almost insensitive to
cloud height whereas both $P$ and its slope decrease with increasing
cloud height. Here we also see that the strength
of the O$_2$A absorption band depends on cloud altitude.
The average length of photon propagation paths through the atmosphere
is shorter for higher clouds; 
typically photons enter the atmosphere at the top, are
scattered in the cloud, and leave the atmosphere. Therefore there is
more molecular absorption when the cloud layer is at a lower
altitude, which can be clearly seen in $I$. For $P$ the amplitude of the
O$_2$A band decreases with increasing cloud height and for very high
clouds $P$ becomes smaller in the O$_2$A band than in the continuum.
 
The third row shows the impact of
effective radius: $I$ is almost invariant and $P$ slightly increases
with increasing particle size. The slope of $P$ slightly decreases with
increasing particle size.

The optical thickness of water clouds is usually larger than 10 so
that the surface is not visible through the clouds. Ice clouds are
optically thinner; for cirrus clouds the optical thickness is
typically about 5 or even smaller and the surface can often 
be seen through the clouds. The bottom panels of Fig.~\ref{fig:ice_cloud} show
simulations with different surfaces: for Lambertian surfaces with
albedos 0.0 and 0.4 and for a green surface. In the simulation with the
black surface the water vapor bands are not visible.
Interestingly these become visible when there is surface reflection. 
Reflection by the surface increases the probability that photons that
are transmitted by the cloud toward the surface are reflected back to
space. Some of these photons are absorbed in the lower atmosphere so
that the absorption bands become visible in the spectrum.

\section{Comparison to observations}
\label{sec:comp_observations}

\subsection{Approximation of Earthshine spectra by weighted sum}
\label{sec:weighted_sum}

\begin{table}[b]
  \centering
  \begin{tabularx}{\hsize}{p{0.4\hsize}cc}
    \hline
    & \multicolumn{2}{c}{Observation date} \\
    Type                       & 25 April 2011 &  10 June 2011 \\ \hline
    Ocean                      &            18 & 46 \\
    Vegetation                 &             7 & 3 \\
    Tundra, shrub, ice, desert &             3 & 1 \\
    Total cloud fraction       &            72 & 50 \\
    Cloud fraction $\tau>$6    &            42 & 27 \\ \hline
    \vspace{0.5ex}
    Model parameters & & \\ \hline
    Ocean BPDF                 & 18        & 46  \\
    Spectral albedo of grass   & 7         & 3   \\
    Albedo 0.4                 & 3         & 1   \\
    Liquid clouds              & 42        & 27  \\
    Ice clouds                 & 30        & 23  \\ \hline 
  \end{tabularx}
  \caption{Top part shows contributions of different types of surface to the
    Earthshine and cloud fractions derived from MODIS data
    \citep[see][Table 1]{sterzik2012}. All fractions are given in percent.
    The bottom part shows the corresponding model parameters that were
    used for the simulations shown in
    Fig.~\ref{fig:model_measurement}.}
  \label{tab:obs_fractions} 
\end{table}
First we approximated the Earthshine observations by the
weighted sum method using the  
fractional surface and cloud cover that have been derived by
\citet{sterzik2012}. The fractions in percent are given in the upper
part of Table~\ref{tab:obs_fractions}. 
The corresponding model parameters are given in the lower part of the
table. We modeled ocean reflection via the BPDF by
\citet{mishchenko1997} with a realistic wind speed of 10~m\,s$^{-1}$
\citep[compare][Fig.~7]{bentamy2003}. We tested the effect of
wind speed on the polarization spectra and found that it is very small
when the full sunglint is in the observation. 
The width of the sun-glint area increases with increasing wind speed but the integral
over it remains approximately the same for all Stokes components.
The difference between
wind speeds of 1~m\,s$^{-1}$ and 10~m\,s$^{-1}$ is less than 0.5\% in P. 
For vegetation we used the spectral albedo of grass as measured by 
\citet{Feister1995}. Other surface types have a very small impact as
their maximum fractional contribution is only 3\% at maximum. We used a
Lambertian albedo of 0.4 as typical for desert to simulate these other surfaces.
The fractional cloud cover
and cloud optical thicknesses were derived from MODIS
observations. We make the simplified assumption that the optically
thicker clouds ($\tau>$6) are liquid water clouds and the thinner
clouds ($\tau<$6) are ice clouds. We further assume that the liquid
water clouds extend in a layer from 2--3~km, have an optical thickness
of 10, and an effective radius of 10~$\mu$m. The ice clouds extend from
10--11~km altitude, have an optical thickness of 1 and an effective
radius of 30~$\mu$m. In all simulations we used the standard
atmosphere for midlatitude summer
\citep{anderson1986}. The phase angle was set to 90\degree.
We obtain the degree of polarization for the ``quasi'' inhomogeneous
planet according to Eq.~\ref{eq:quasi_inhom}. 
\begin{figure}[h]
  \vspace*{2mm}
  \begin{center}
    \includegraphics[width=1.0\hsize]{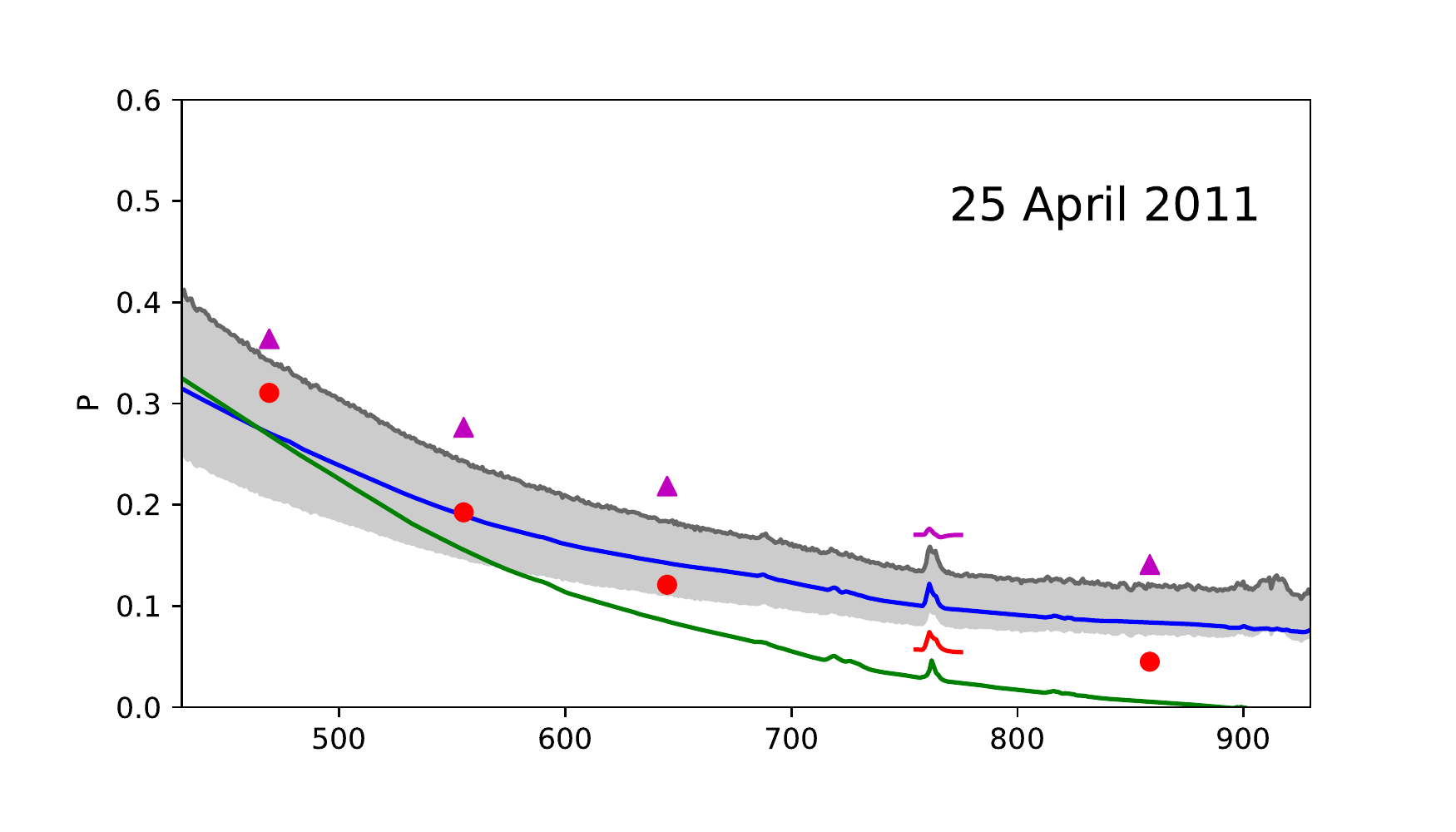}\\
    \includegraphics[width=1.0\hsize]{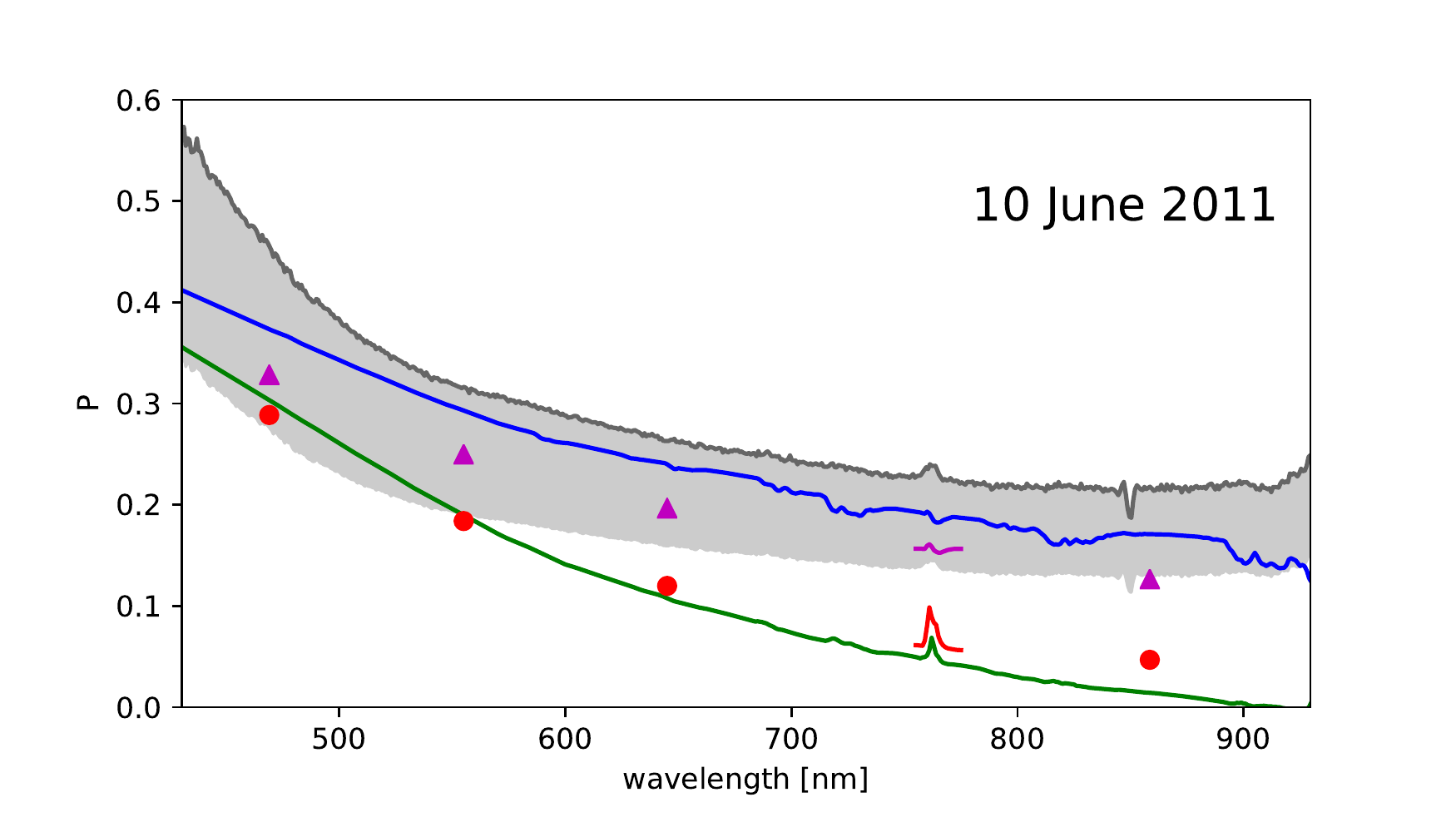}
  \end{center}  
  \caption{Observations by FORS against MYSTIC simulations (see text for
    details). The blue lines correspond to simulations for homogeneous
    planets, which are averaged using a weighted sum approach (see
    Section~\ref{sec:weighted_sum}). The red
    circles correspond to simulations for a realistic Earth model
    including 3D cloud fields and a 2D surface albedo map derived from
    MODIS data (see Section~\ref{sec:2Dsurface}). 
    The purple triangles indicate a planet with the same 3D
    cloud fields, but the surface is completely covered with ocean and
    simulated using a BPDF (see Section~\ref{sec:ocean_bpdf}). 
    The gray area shows the uncertainty range due to the unknown moon depolarization.
    The green line shows the best fit, which was obtained 
    using the dataset by \citet{stam2008} without ice clouds 
    \citep[see][]{sterzik2012}. The top panel represents the observation
    on 25 April 2011 and the bottom panel 10 June 2011.}
  \label{fig:model_measurement}
\end{figure}

Figure~\ref{fig:model_measurement} shows simulated (blue) versus observed
(gray) spectra of the degree of polarization. The gray area shows the
uncertainty of the observed degree of polarization due to the
uncertain depolarization of the moon surface (see Section~\ref{sec:observation}). 
The green lines show the best fit that
could be obtained using the dataset by \citet{stam2008}, which includes
various surface types and liquid water clouds, but no ice water
clouds. Further the fit does not include a realistic BPDF to simulate
polarized reflection at the ocean surface. 
Both scattering by high ice water clouds and reflection at water
surfaces produce a higher degree of polarization in the red part of the
spectrum and decrease the spectral slope,
therefore we obtain a much better match with the observation.

Indeed, the simulations for 25 April fit the observations
very well within the quite large uncertainty due to the uncertain moon
depolarization. The magnitude and slope of the spectrum of
the degree of polarization are very similar.  
The simulations for 10 June are also within the uncertainty range of
the observation. However, the spectral slope is different. The
very steep decrease from 400--500~nm followed by the rather flat
spectrum from 500--900~nm could not be simulated and
the increase of the degree of
polarization above 900~nm cannot be reproduced by the model. 
Further, in the O$_2$A band region, the degree of polarization is decreased
compared to the continuum in the model whereas it is increased in the
observation. The decrease can be explained by the very high ocean
fraction that is modeled using the BPDF.
Whereas Lambertian surface reflection is unpolarized, 
the ocean reflection is highly polarized in the sunglint region. 
For Lambertian reflection $Q$ and $U$ are zero for radiation that is
reflected, hence only $I$ is reduced by absorption so that the degree
of polarization is higher in the O$_2$A band than in the continuum.
Radiation reflected at the ocean is polarized, hence $I$, $Q$ and $U$
are decreased by absorption and depending on the reflectance function,
the polarization in the  O$_2$A band can be higher or lower compared
to the continuum. Unpublished Earthshine spectra of the O$_2$A band also
exhibit large variability, including the possibility of a
smaller band polarization compared to the continuum \citep{sterzik2017}.
Our result shows that the weighted sum approach is
not accurate enough to explain the shape of the O$_2$A band.

We know from the results by \citet{karalidi2012a} that the
approximation by a weighted sum of results for homogeneous planets is
not accurate. However, the results shown in this section already
indicate that the observations are realistic, in particular we now
understand the high degree of polarization in the continuum of 
the red part of the spectrum.

\subsection{Realistic simulation with 3D cloud fields}

\subsubsection{Simulation with 2D Lambertian surface}
\label{sec:2Dsurface}

We simulated the Earthshine spectra for a realistic Earth
atmosphere including 3D clouds from the ECMWF model as
in Section~\ref{sec:earth_pics}. We used cloud data from the operational
forecast system IFS. For the observation on 25 April 2011, 9:00 UTC,
we took the 9 hour forecast from midnight (25 April 2011, 0:00
UTC) and for the observation on 10 June, 1:00 UTC, we take the 12 hour
forecast from 9 June 2011, 12:00 UTC. 
As before we used constant effective radii of 10~$\mu$m for liquid
water clouds and 30~$\mu$m for ice water clouds. Further
we multiplied the liquid/ice water contents with the sub-grid cloud
cover to obtain correct water contents for each grid cell. 

To define the geometrical setup for the MYSTIC simulation, 
we provide the exact 
position of the moon with respect to the Earth (latitude and
longitude, distance) and the position of the sun.
Table~\ref{tab:sun_moon_pos} shows the positions that were used for the
simulations, these setups correspond to the configurations on
25 April 2011 and on 10 June 2011, when Earthshine spectra were observed.
\begin{table}
  \centering
  \begin{tabular}[h]{l|ll}
    \hline
    & 25 April 2011, 9:00 UTC & 10 June 2011, 1:00 UTC \\ \hline
    sun & 13\degree N, 45\degree E & 23\degree N, 165\degree E \\
    moon & 15\degree S, 39\degree W & 5\degree S, 95\degree W \\ \hline 
  \end{tabular}
  \caption{Positions of sun and moon used for the simulations. }
  \label{tab:sun_moon_pos}
\end{table}
Further we define the field of view of the sensor so that it
includes the full Earth and set the spatial resolution to
500$\times$500 pixels.
 
We used the 2D surface albedo map derived from MODIS data (see
Sec.~\ref{sec:earth_pics}). The surface is thus treated as Lambertian
reflector with different albedos, where the albedo of the ocean is
very small. Ocean reflection is not modeled using the BPDF in
this section.  The molecular atmosphere (US standard) is constant over
the full planet and
molecular absorption is calculated with REPTRAN.

As mentioned before, we cannot include spectrally dependent cloud
optical properties when the ALIS method is used.
Therefore
we performed only monochromatic calculations for center wavelengths
of MODIS channels (469~nm, 555~nm, 645~nm, and 858.5~nm), where
2D surface albedo data derived from MODIS are available.

\begin{figure*}[t!]
  \begin{center}
    \includegraphics[width=1.0\hsize]{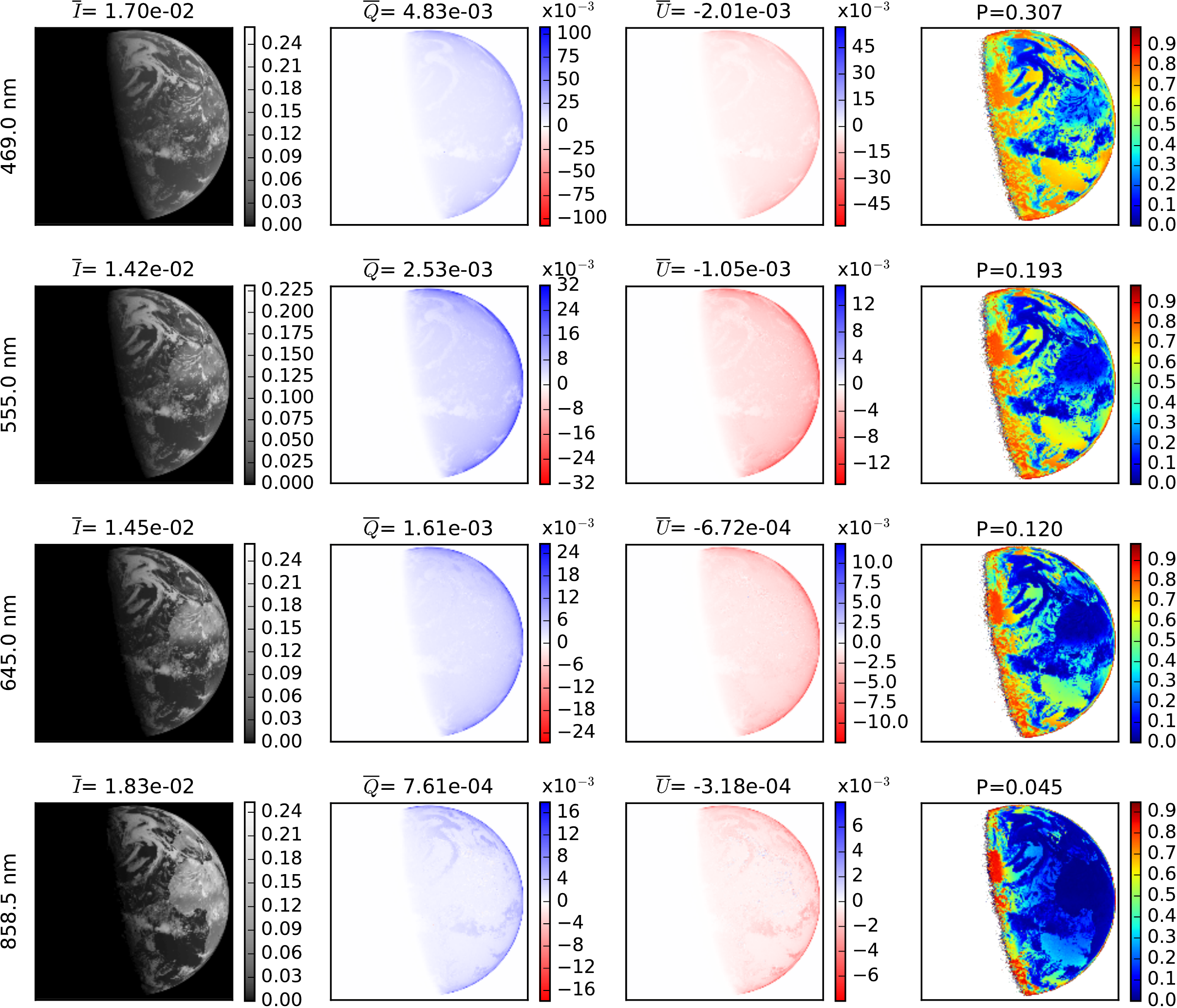}
  \end{center}
  \caption{Simulations of the Earth as seen by
    the moon on 25 April 2011 at 9:00 UTC. The cloud fields (liquid
    and ice clouds) are taken from the ECMWF IFS weather forecast
    model. The surface is included as a 2D albedo map
    derived from MODIS observations. The rows correspond to the
    central wavelengths of MODIS channels. The left plots show the
    radiance $I$, the middle plots show the linear polarization
    components $Q$ and $U,$ and the right plot shows the degree of
    polarization $P$. The numbers on top of the figures indicate
    the averages for $I$, $Q,$ and $U$ and the degree of polarization
    of the complete image.}
  \label{fig:real_sim_wvls}
\end{figure*}
Figure~\ref{fig:real_sim_wvls} shows the results for all simulated
wavelengths. The left images correspond to the intensity (Stokes
component $I$ normalized to incoming solar irradiance). At 469~nm we
hardly see the surface because land surfaces reflect only little
radiation at short wavelengths and the ocean albedo is close to 0. The
surface looks brightest at 858.5~nm because at this wavelength 
the land surface albedo is very high, whereas the clear-sky
parts appear very dark because there is only very little Rayleigh
scattering. In all images of $Q_i$ and $U_i$ the land surface is not
visible because we assume a Lambertian non-polarizing surface. At the
shorter wavelengths 469~nm, 555~nm, and 645~nm 
the clouds depolarize, i.e., $Q_i$ and $U_i$ have
smaller absolute values above clouds. The depolarization by clouds can
also be seen in the image of the degree of polarization $P_i$. 
For 858.5~nm clouds polarize
more strongly than the Rayleigh background, thus they appear darker in the
$Q_i$ and $U_i$ images. The degree of polarization $P_i$ is generally very
small and close to 0 above the bright land surface. 
The numbers above the images are mean values of the Stokes vector
components $\overline{I}=\sum{I_i}/N$, $\overline{Q}=\sum{Q_i}/N$ and
$\overline{U}=\sum{U_i}/N$, where $N=500^2$ is the number of pixels.
The degree of polarization
$P=\sqrt{\overline{Q}^2+\overline{U}^2}/\overline{I}$ is written above
the image of $P_i$. The value $P$ decreases
from about 0.3 at 469~nm to 0.045 at 858.5~nm. 
These results are indicated as red circles in
Figure~\ref{fig:model_measurement}. The value at 858.5~nm is clearly
lower than the measurement; all other results are within the gray area
and hence match the observation within the range of uncertainty. 
Using the same model setup we also calculate the region about the
O$_2$A band from 755~nm to 775~nm using ALIS
with constant scattering coefficients over the spectral
range. We find a relatively strong O$_2$A-band polarization compared
to the continuum, i.e., stronger than in the observation. 

The same simulations were performed for 10 June 2011 and the
resulting values of the degree of polarization are also included in
Figure~\ref{fig:model_measurement}. For this observation day the results
are far below the observation for 645~nm and 858.5~nm. The degree of
polarization is
very similar to the results for the April observation, which means
that the globally averaged cloud distributions from the ECMWF model are similar for
the two observation times in April and June. In particular the clouds do not explain
the observed difference in degree of polarization of almost 10\% in
the red part of the spectrum. 

\begin{figure*}[t!]
  \begin{center}
    \includegraphics[width=0.4\hsize]{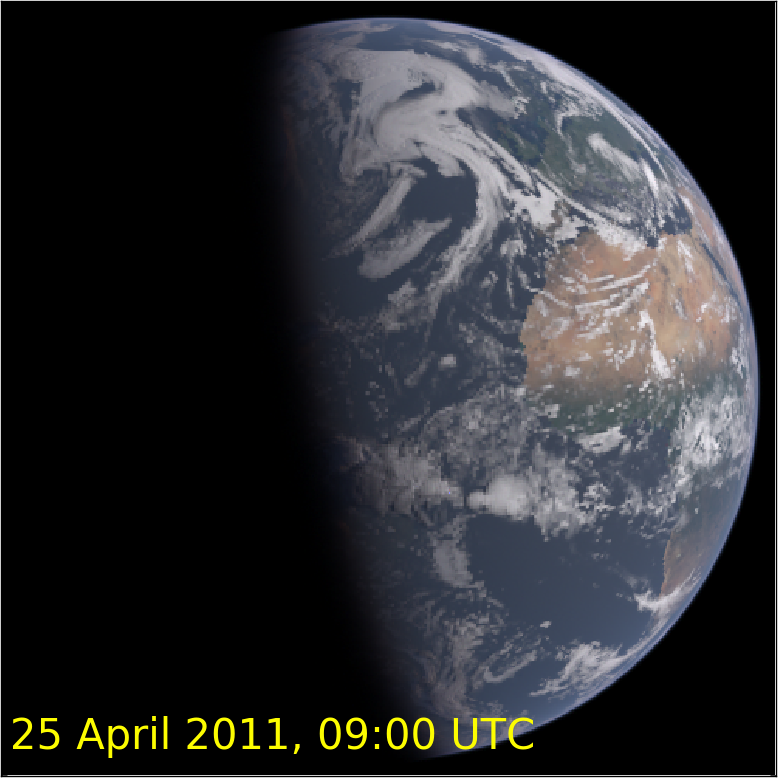} \hspace{2cm}
    \includegraphics[width=0.4\hsize]{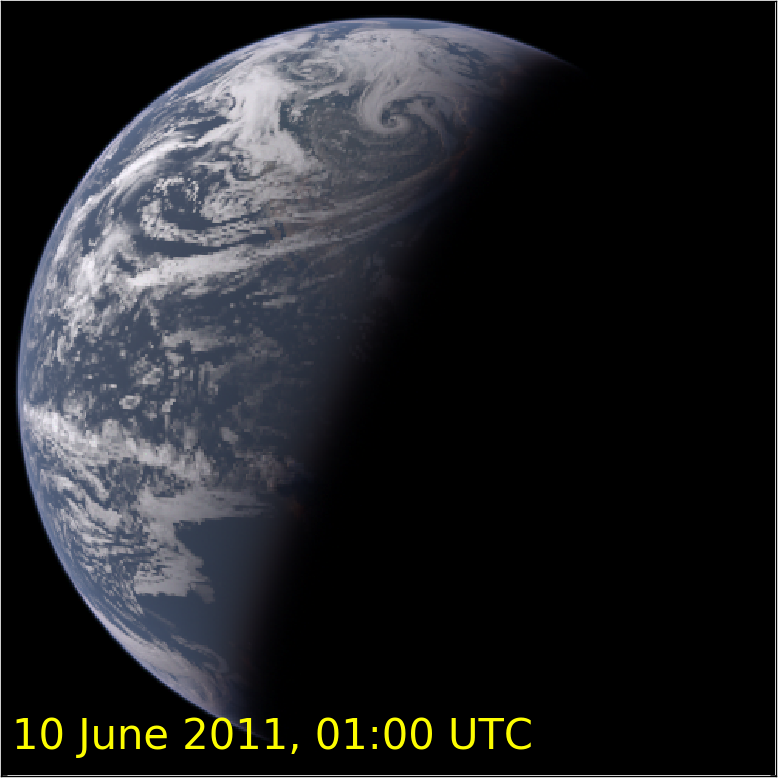}
  \end{center}  
  \caption{Simulations of the Earth as seen by
    the moon. The figure shows a true color composite; red
    corresponds to 645~nm, green to 555~nm, and blue to
    469~nm.}
  \label{fig:real_simulation}
\end{figure*}
\begin{figure*}[htbp]
  \begin{center}
    \includegraphics[width=1.0\hsize]{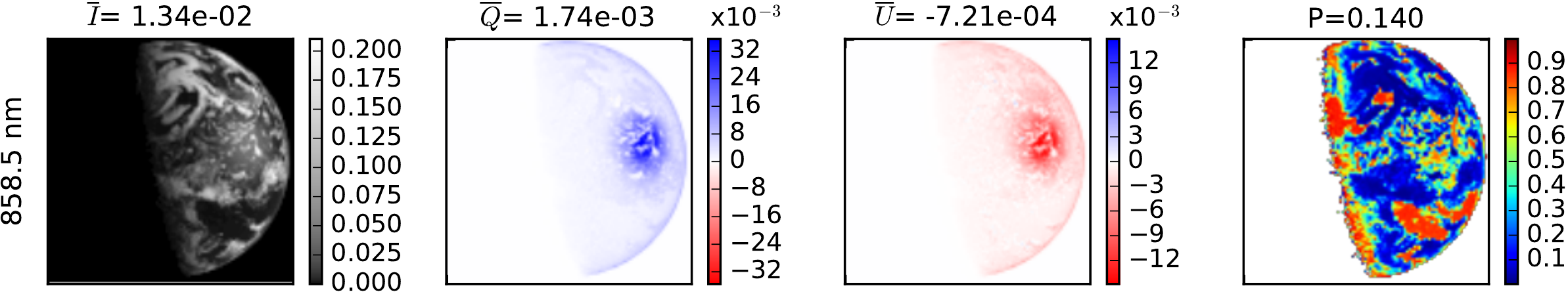}\\
    \includegraphics[width=1.0\hsize]{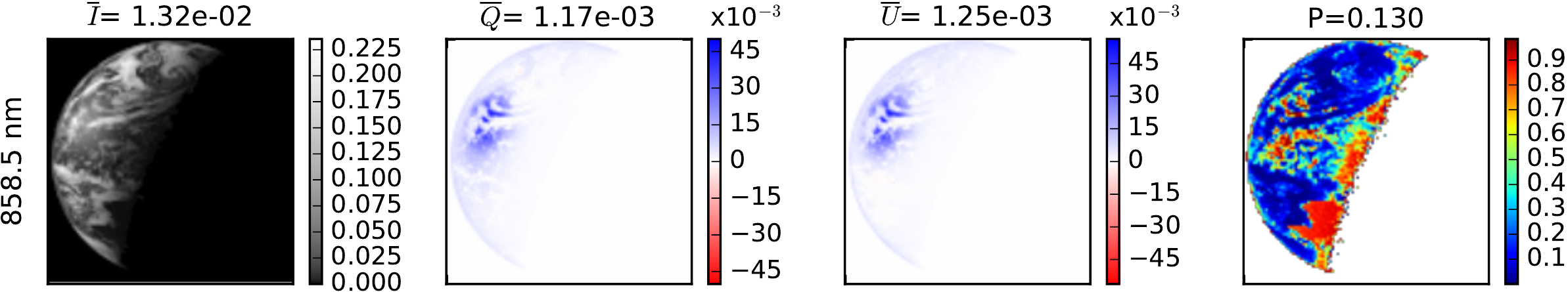}
  \end{center}
  \caption{Simulations for a planet covered with liquid water. The
    geometrical setup corresponds to the two measurement dates, 25
    April 2011 (top) and 10 June 2011 (bottom). The ocean surface
    reflection is considered by a realistic bi-directional polarized
    distribution function (BPDF) \citep{mishchenko1997}. The sunglint
    is clearly visible in the images of the  Stokes components $I_i$, 
    $Q_i$ and $U_i$.}
  \label{fig:bpdf_simulation}
\end{figure*}
Figure~\ref{fig:real_simulation} shows a true color composite, where red
corresponds to 645~nm, green to 555~nm, and blue to 469~nm. The
results look very similar to satellite images of the Earth, and indeed
clouds and surface look very realistic. But quantitatively the degree
of polarization is too low, especially in the red part of the
spectrum for 10 June 2011. 

One important aspect is still missing in these simulations: the contribution to the observed polarization due to the scattering of
the solar radiation by the surface.
Whether or not this explains the
remaining differences between simulation and observation is
investigated in the next section.

\subsubsection{Polarization by ocean}
\label{sec:ocean_bpdf}
Currently our
model does not allow us to combine the ocean BPDF with a Lambertian
land surface. In order to test the impact of polarization by
reflection at the water surface, we performed simulations for a planet
completely covered with water. We used a realistic ocean BPDF
\citep{mishchenko1997}, which
includes also shadowing effects by ocean waves. We assumed a constant
wind speed of 10~m\,s$^{-1}$ and performed monochromatic calculations at
469\,nm, 555\,nm, 645\,nm, and 858.5\,nm. The spatial resolution of the
images is set to 100$\times$100 pixels.

Figure~\ref{fig:bpdf_simulation} shows the results at 858.5~nm. The
top panels are for 25~April~2011 and the bottom panels for
10~June~2011. The images of $I_i$ show the sunglint as a bright
area. In
the images of $Q_i$ and $U_i$, the high polarization in the sunglint
region is clearly seen. The pixel-average $\overline{Q}$ and
$\overline{U}$
are significantly higher than in the simulations for the 2D-albedo map
from MODIS data, and this yields an increase of about 10\% in degree of polarization $P$.
When we compare Fig.~\ref{fig:real_simulation} and
Fig.~\ref{fig:bpdf_simulation} we see that 
the sunglint on 25 April 2011 at 9:00 UTC
would be located on the Sahara desert,
thus at this time we do not expect much polarization from radiation reflected by
the ocean. On 10 June at 1:00 UTC the sunglint is located in the middle of
the Pacific ocean, so we can expect to see it clearly in the observation.   
The results of all simulations for a completely water covered planet
are included as magenta triangles in
Fig.~\ref{fig:model_measurement}.

On 25 April the predicted degree
of polarization for the planet completely covered with water
is higher than the observation (purple triangles). For this day the
simulation with the 2D Lambertian surface albedo (red circles) is more
realistic than the simulation for a water covered planet.
The remaining difference between observation and simulation (red
circles) can be
explained by the completely missing polarization contribution 
by the surface in the simulation with Lambertian surface. 

On 10 June the simulations for the water covered
planet (magenta triangles) are more realistic and indeed they lie in
the gray area corresponding to the uncertainty range of the
observation. 

The results clearly show that the difference between the two
observed spectra of about 10\% in the red part of the spectrum is due to
the polarized reflection of the ocean surface, in particular in the
sunglint region.

\subsection{Computational times}
\label{sec:cpu_times}

\begin{table}[b]
  \centering
  \begin{tabularx}{\hsize}{p{0.5\hsize}cccc}
    \hline
    Homogeneous planets & $\overline{P}$ & $\overline{\sigma(P)/P}$ & time [s] \\ \hline
    BPDF (Tsang) & 0.84 & 0.010 & 477 \\
    Albedo 0.4 & 0.18  & 0.015 & 490 \\
    Spectral albedo of grass & 0.46 & 0.010 & 480 \\ 
    Liquid cloud ($\tau$=10.0) & 0.06 & 0.088 & 802\\
    Ice cloud ($\tau$=1.0) & 0.21 & 0.025 & 687 \\[1ex] \hline
    Inhomogeneous planets & $P$ & $\sigma(P)/P$ & time [s] \\ \hline
    ECMWF clouds \& MODIS albedo & 0.18 & 0.006 & 29 \\
    ECMWF clouds \& BPDF  & 0.25 & 0.005 & 28 \\ \hline 
  \end{tabularx}
  \caption{Computational time on one CPU (Intel(R) Xeon(R) CPU E5-2630 v4 @ 2.20GHz)
    for the full spectra (1000 wavelengths) of the degree of polarization 
    and corresponding spectrally averaged relative
    standard deviations $\sigma$. 
    The simulations were performed with $N_{\rm ph}$=10$^{5}$ photons. 
    The computational time is proportional to $N_{\rm ph}$ and the 
    corresponding standard deviation $\sigma$ is proportional to
    $N_{\rm ph}^{-1/2}$. 
    The numbers for the scenarios with ECMWF clouds correspond to
    monochromatic calculations at 555~nm for 10 June 2011.
   }
   \label{tab:cpu_times}
\end{table}
The computational times for the cases shown in
Fig.~\ref{fig:model_measurement} are summarized in
Table~\ref{tab:cpu_times}. In the Monte Carlo codes the computational time
is proportional to the number of photons $N_{\rm ph}$ used for the simulation and
the standard deviation is proportional to $N_{\rm ph}^{-1/2}$. All
results shown in the table are for $N_{\rm ph}$=10$^{5}$.
For the homogeneous planet simulations we used the absorption lines
importance sampling method, ALIS, and the given times are for full spectra
with 1000 grid points. The clear-sky simulations take less than 
10~minutes and simulations with clouds take less than 15~minutes. 
For the inhomogeneous planets including the 3D cloud fields the 
computational time corresponds to monochromatic calculations. 
These simulations (with 2D albedo map or BPDF) take about 0.5 minutes.
The images require significantly
more computational time. If we want to have the same accuracy for each pixel
we have to multiply the given computational 
time by the number of pixels; for example, for
the images with 100$\times$100 pixels the computational time is
increased by a factor of 10$^{4}$.

\section{Conclusions}
\label{sec:conclusions}

We have presented a novel approach to simulate the polarized radiation
scattered by Earth as seen from space, which is of general
significance also for future polarimetric observations of Earth-like
exoplanets. The approach has been implemented in
the Monte Carlo code MYSTIC, which allows us
to compute the intensity and degree of polarization
with high accuracy and high spectral resolution. We validated the
outputs of the code by comparing its predictions with the data
tabulated by Stam (2008) for homogeneous planets with different
surfaces, and we consistently found very good agreement. Then
we investigated the impact of various atmospheric components:
aerosols, water clouds, and ice clouds, as well as that of various
kinds of surface. We found, for instance, that the presence of high
ice clouds in the atmosphere increases the degree of polarization in
the red part of the spectrum and that the polarization due to the
scattering of the light from the ocean has a large impact if the
sunglint over ocean is visible to the observer. The
potential to detect enhanced reflectivity from an ocean glint and its
effect on polarization in the phase curve of an Earth-like exoplanet
has been previously advocated by \citet{williams2008}.

We used our code to try to interpret the Earthshine optical
spectropolarimetric observations obtained in April and June 2011 with
the FORS instrument of the ESO VLT by Sterzik et al. (2012). In
particular we were interested in
explaining the high value of the polarization observed at longer
wavelengths that could not be interpreted in the previous modeling
attempts. 
We performed simulations considering realistic 3D fields of
liquid water and ice water clouds, which we obtained from the ECMWF
operational weather forecast model at the times of the FORS
observations, and 2D albedo maps of the planet surface, which were
derived from MODIS observations.
This way we obtained very similar results for the two specific
observing epochs, fitting the April data well, but not the
observations obtained in June, which exhibited a degree of
polarization almost 10\,\% higher in the red part of the spectrum than
those obtained in April.
We then performed a simulation with the same
cloud fields, but for a planet surface completely covered by water, using a realistic bi-directional polarized distribution function. By
so doing, we substantially improved the fit to the observations
obtained in June. By simulating images of the Earth as seen by the Moon we
found that in June, the position of the sunglint is above the Pacific ocean,
whereas in April the position would be above the Sahara, and thus the
observation in April does not include the sunglint.
Our modeling results show that the degree of polarization in the
red part of the spectrum increases by about 10\% when the sunglint
is observed; this increase corresponds to the observed difference between the
two observation epochs.

In conclusion, compared to previous modeling
results, the inclusion of ice clouds allowed us a substantial
improvement of the fit to the data obtained in both observing
epochs. We then showed that the enhanced polarization fraction
caused by the sunglint over ocean can be clearly 
identified in the red part of the Earthshine spectra obtained in June.
This finding may be of 
significance in the interpretation of future polarimetric measurements
of exosolar planets.

To further improve the modeling of the polarization properties of the
radiation scattered by Earth, the following steps are needed:
\begin{itemize}
\item Allowing the simultaneous inclusion of BPDF and Lambertian surfaces
\item Implementation of BPDFs for land surfaces
\item Inclusion of 3D aerosol data
\end{itemize}

A major difficulty in the interpretation of the Earthshine
observations is the uncertain contribution of the depolarization of
the moon surface. In that respect, a fully reliable modeling of the
light scattered by Earth could be actually used to measure the
depolarization matrix of the moon surface from Earthshine measurements.

Future applications of our model may include the interpretation
of spectropolarimetric observations in the near-infrared spectral range
\citep[e.g.,][]{miles-paez2014} and the simulation of
images of the degree of polarization of other planets 
of the solar system, for example,
using the Jupiter maps observed by 
\citet{mclean2016}.

\begin{acknowledgements}
Based on observations made with ESO Telescopes at the La Silla Paranal
Observatory under program 87.C-0040.
We thank the anonymous reviewer for helpful comments and suggestions
that helped us to improve our manuscript. 

\end{acknowledgements}

\bibliography{./literature}
\bibliographystyle{aa}


\end{document}